\documentclass[12pt]{JHEP3}
\usepackage{graphicx}
\usepackage{epsf,
shadow,pifont}
\usepackage{amsmath,epsfig}
\usepackage{amssymb,amsfonts}
\usepackage{latexsym}
\usepackage{slashed}
\usepackage{epsfig}
\usepackage{xcolor}
\usepackage{subfigure}
\usepackage{subfig}
\usepackage{comment}
\let\normalcolor\relax
\usepackage[vcentermath,enableskew]{youngtab}
\def\hri#1#2{\href{http://arxiv.org/abs/#1}{[ArXiv:#1]#2}}
\def\hre#1#2{\href{http://arxiv.org/abs/#1/#2}{[ArXiv:#1/#2]}}

\relax
\renewcommand{\theequation}{\arabic{section}.\arabic{equation}}

\def\be{\begin{equation}}
\def\ee{\end{equation}}
\def\qslash{{q\hspace{-5.3pt}/}}

\def\kslash{{k\hspace{-5.3pt}/}}

\def\eslash{{\epsilon\hspace{-5.3pt}/}}

\newcommand{\bear}{\begin{eqnarray}}
\newcommand{\bea}{\begin{eqnarray}}
\newcommand{\eear}{\end{eqnarray}}
\newcommand{\eea}{\end{eqnarray}}
\def\hre#1#2{\href{http://arxiv.org/abs/#1/#2}{[ArXiv:#1/#2]}}

\newbox\pippobox

\def\II{\relax{\rm I\kern-.18em I}}

\def\e{\epsilon}
\def\m{\mu}
\def\n{\nu}
\def\r{\rho}
\def\s{\sigma}
\def\pa{\partial}

\def\sp{\;\;\;,\;\;\;}

\def\tr{\ensuremath{\mathrm{Tr}}}

\def\l{\lambda}
\def\g{\gamma}
\def\d{\delta}
\def\half{\frac{1}{2}}

\def\AA{{\cal A}}
\def\BB{{\cal B}}
\def\CC{{\cal C}}

\def\II{{\cal I}}

\def\LL{{\cal L}}

\def\nn{\nonumber}
\def\trule{\noalign{\hrule}}

\renewcommand{\>}{\rangle}

\newcommand{\red}[1]{{\color{red} #1 \color{black}}}
\newcommand{\blue}[1]{{\color{blue} #1 \color{black}}}

\newcommand{\EK}[1]{{\red{EK: #1}}} 
\newcommand{\PB}[1]{{\blue{PB: #1}}} 

\title{Emergent/Composite axions}

\author{\large
P.~Anastasopoulos$^{1}$, 
P.~Betzios $^{2}$, 
M.~Bianchi$^{3}$,

D.~Consoli$^{1,3}$, 
E.~Kiritsis$^{2,4}$
 ~\\
{$^1$ Mathematical Physics Group, University of Vienna, Boltzmanngasse 5 1090 Vienna, Austria.}
 ~\\
{$^2$ \href{http://hep.physics.uoc.gr/}{Crete Center for Theoretical Physics}, Department of Physics, University of Crete,
Herakleio, Greece.} ~\\
{$^3$ Dipartimento  di  Fisica,  Universit\`a  di  Roma  ``Tor  Vergata''  \& \\ I.N.F.N.  Sezione  di  Roma  ``Tor  Vergata'', Via della Ricerca Scientifica, 00133 Roma, Italy.}
 ~\\
{$^4$ \href{http://www.apc.univ-paris7.fr}{APC, Universit\'e Paris 7}, CNRS/IN2P3, CEA/IRFU, Obs. de Paris, Sorbonne Paris Cit\'e, B\^atiment Condorcet, F-75205, Paris Cedex 13, France (UMR du CNRS 7164).}}

\preprint{CCTP-2018-13\\
ITCP-2018-10\\
UWThPh2018-28}

\abstract{Hidden theories coupled to the SM may provide emergent axions, that are composites/bound-states of the hidden fields. This is motivated by paradigms emerging from the AdS/CFT correspondence but it is a more general phenomenon. We explore the general setup and find that UV-sourced interactions of instanton densities give rise to emergent axions in the IR. We study the general properties of such axions and argue that they are generically different from both fundamental and composite axions that have been studied so far.
      }


\begin{document}

\section{Introduction and results}


The notion of an axion goes back to the seminal work of Peccei and Quinn, \cite{PQ} who introduced it in order to provide a natural solution to the strong CP-problem.
The original theory was not renormalizable because of the axion coupling to the QCD-instanton density \cite{Witten:1979vv, Veneziano:1979ec}.
Quickly afterwards,  renormalizable theories of axions were constructed \cite{Kim}-\cite{Zhitnitsky}, that also made the axion weakly interacting so as to avoid direct experimental constraints.
Such axions came under the name of ``invisible" axions and their couplings to matter have been determined using anomalies and the chiral Lagrangian \cite{Kaplan}-\cite{DiVecchia:2017xpu}.
They  have become over the years the objects of both theoretical and experimental scrutiny, especially as they are prime candidates for the dark matter of the universe, but also candidates for the inflaton, \cite{Kim2}-\cite{IR}.

Axions as scalar fields in an effective field theory are special. They always have a perturbative shift symmetry.
It is this perturbative symmetry and the fact that they couple to instanton densities that provides a definition of what is an axion.
The issue of the symmetry is however subtle: in all the cases we know and which make sense as QFTs, such a symmetry is broken to a discrete symmetry (at best) due to non-perturbative effects. In weakly-coupled theories, such effects are associated to instantons.

In the canonical case of the QCD axion, such instanton-related effects are responsible for giving a mass and a potential to the axion (see \cite{Villa} for a recent exposition).

Axions are ubiquitous in string theory (see \cite{SW} for a review and \cite{Axi} for an effective theory discussion).
They appear in two forms. Either as pristine massless scalars as the Ramond-Ramond (RR) axion of type IIB string theory.
Or as internal components of antisymmetric form gauge fields upon compactification, as well as off-diagonal components of the metric. Both types unite in that they are generalized gauge fields of string theory and therefore have accompanying gauge symmetries \cite{pol,k}.
It is these symmetries that provide the perturbative Peccei-Quinn (PQ) symmetries in String theory \cite{s}.

It is also the case that continuous shift symmetries, that appear as would-be global symmetries in string theory, are broken by non-perturbative effects, at best to discrete ones, in agreement with the postulate and  evidence that there are no global continuous symmetries in string theory.
The argument is quite general\footnote{In fact {\it mutatis mutandis} it applies to NS-NS axions that couple to world-sheet or NS5-brane instantons.}: RR axions couple to the world volume of D-branes as shown in general in \cite{pol2}. The same D-branes, wrapped around an appropriate Euclidean internal cycle, provide instanton effects in string theory \cite{BBS}-\cite{revi}. The nature of these effects depends on the amount of supersymmetry. In the case of maximal supersymmetry they do not generate a potential, but affect higher derivative terms like the $R^4$ corrections, \cite{gg,kp}, that are reproduced by the AdS/CFT correspondence \cite{Bianchi:1998nk, Bianchi:2007ft, Bianchi:2009ij}.
With less or no supersymmetry, they generate a (super)potential for the axions \cite{Bianchi}-\cite{Bianchi:2015cta} as it happens for QCD.
In both cases, the end result is the same: the would-be global symmetry is broken to a discrete subgroup. Moreover, even this remaining discrete symmetry is gauged in string theory, because the original shift symmetry was a remnant of a gauge symmetry.
There are subtleties in the above that arise when the axionic symmetries are coupled to anomalous U(1)'s \cite{Bianchi}-\cite{Bianchi:2015cta}, but do not change the final result.

In all the above descriptions, the axion fields are fundamental fields in the theory, be it a string theory or a QFT.
In this paper, we would like to discuss a novel type of axion, the composite or emergent axion.
The composite axion idea is not new, \cite{Kim1}-\cite{Wil}.
In all its realizations, it involved a strongly-coupled and confining theory, the Peccei-Quinn symmetry is a symmetry acting on fermions and the role of the axion was played by one of the axi-pions that remains massless in perturbation theory.
The characteristic energy scale, $f$, for the composite axion was the strong coupling scale of the axi-gauge theory, $f\sim \Lambda_a$, that is responsible for the interaction generating the bound state, since
the mass of the axion  is generated by QCD effects. Therefore, to generate a reasonable composite axion that is ``invisible" one needs $\Lambda_a\gg \Lambda_{QCD}$, \cite{Kim2}. As somewhat different approach was used in a RS setup in \cite{Die}.

It was also realized that there are phenomenological differences in such realizations compared to fundamental axions. As an example, in simple models, the axion does not couple to the lepton sector.

{Our setup is general and will describe composite axions arising from ``hidden sectors".
The hidden sector theory couples to the SM at some scale $M$. We will assume for most of the paper that $M$ is much larger than SM scales, but in some cases we will also consider what happens when $M$ is of the order of SM scales or much smaller.

Our main goal however in this paper is to specifically analyze the case where the hidden theory is a holographic theory (ie. a large-N, strongly coupled gauge theory). In this case, we will call the composite axions ``emergent".
As we shall see, such axions are novel, both in terms of their identification
(they will be instanton densities instead of $\eta'$s) and their properties.
Moreover, in the holographic case, despite the strong coupling, we will have  the extra advantage of perturbativity while the emergent axion that is generated is tightly bound.

As we shall see in section \ref{PhenoConsiderations} instanton densities in the holographic hidden theory can play the role of QCD axions solving the strong CP problem. Axions emerging from other hidden theories cannot be QCD axions but can be axion-light particles playing a role in the early universe cosmology.
In most of the paper we will consider a generic hidden sector theory
and we will specialize to the holographic case in section \ref{Hologaxion}.

In the context of a holographic hidden theory, the emergent axion is an avatar of the generation of emergent gravity at the same time, \cite{1}.
The hidden-sector theory and the SM are coupled via interactions that are irrelevant in the IR. This fact ensures that one can have an invisible emergent axion even if  the strong coupling scale of the hidden gauge theory
$\Lambda_{h}\ll \Lambda_{QCD}$.}

The emergent axions we discuss {\em do not rely} on fermionic Peccei-Quinn like symmetries, but rather on the approximate Peccei-Quinn symmetry associated to instanton densities in gauge theories.
We would like to expand a bit upon this as it seems puzzling at first.
There are two important ingredients  for any solution to the strong CP-problem. The first is a dynamical field (we call it an axion) that couples linearly to the associated QCD instanton density. This guarantees that its vev can shift the UV $\theta$-angle of QCD. Typically this is implemented by an approximate shift symmetry that implies that constant shifts in $\theta$ can be absorbed in shifts in the dynamical field. This approximate shift symmetry is the PQ symmetry.

The second is that the effective potential for the dynamical field and the $\theta$-angle has a global minimum at zero value. This guarantees that dynamically, strong CP violation is absent.
It is always the case that the would be PQ symmetry is violated by non-perturbative effects (instantons, if  at weak coupling).
This breaks the PQ symmetry non-perturbatively and gives also a mass to the axion. The phenomenological viability then of the setup relies on the values for the strength of the axion interactions and its mass.
Therefore, even if the two above conditions are met, it is not a given that an axion can solve the strong CP problem. This relies on more detailed dynamical issues.

There are two type of generic (approximate) symmetries that can play the role of a PQ symmetry. The first used in an ubiquitous fashion is a U(1) symmetry acting on fermion fields. It is a symmetry in perturbation theory but it is typically violated by a quantum (triangle) anomaly. This does not violate charge conservation in perturbation theory, but non-perturbative effects break it, and give the axion a mass.

What we advocate here is that there is another symmetry that can play the role of a PQ symmetry. This is the perturbative shift symmetry that is associated with the instanton density operator. This symmetry is exact in perturbation theory, but it is violated non-perturbatively by instanton effects. This again  gives a mass to the axion, similarly to the previous case.
It is clear from string theory (and the prime paradigm of the AdS/CFT correspondence) that QFT instanton densities are associated to string theory axions that have shift symmetries in perturbation theory and which are violated by string instantons.

In the former case, the axion is some combination of $\eta'$s while in the latter it is the $0^{+-}$ glueball. The general case involved various combinations of the two, as they may also be mixing among them.
It should be stressed though, that we may have cases without any $\eta'$ and we can still have an axion.

As mentioned, the paradigm for such a type of composite/emergent axion is already in the AdS/CFT correspondence.
In the holographic context, the (gauge singlet) bulk fields of the dual string theory are thought of as composites made out of the generalized gluons of the (hidden) quantum field theory. In the standard examples, this theory is a CFT, and the notion of a bound state as a ``particle" is somewhat imprecise. One can make it however more particle-like by breaking conformal invariance and producing a non-conformal strongly coupled theory with a non-trivial characteristic scale.

We start by briefly describing some of the ideas in \cite{1}. Their purpose was to describe the SM of particle physics coupled to (semiclassical) gravity (and maybe other interactions) as emerging from four-dimensional QFT's, one of which is the SM.
For simplicity, we assume that beyond the SM there is a single hidden QFT that is strongly coupled and at large N.

The hidden theory will be, among other things, the source of gravity for the SM fields.
The reason that we require large N and strong coupling from the hidden theory is that it should generate semiclassical gravity according to the dictums of the holographic correspondence.

The key point is the coupling of the hidden theory to the SM.
From some general considerations that are detailed in \cite{1}, we would like to couple the two theories in the UV. Modulo subtleties, the only UV complete way of doing this is via a messenger sector that is composed of bifundamental fields, charged under both the hidden gauge group and the SM gauge group.
The messenger fields have large masses, $M$, much above all the mass scales of the SM.
Integrating them out, we obtain the effective action for the low-energy theory
that involves the SM and the hidden theory coupled by double (or multiple) trace interactions, that are all irrelevant\footnote{There is an exception to this, \cite{quiros}, but we shall not consider it in this paper.}.
This is the reason that the novel interactions that are triggered between the two theories are weak at low energies.

Generically, upon integrating out the ``invisible" hidden theory, the inter-theory high-energy interactions can be resolved by effective, low-energy dynamical fields coupled to the standard model.
One of the them, associated to the hidden stress tensor, appears as a dynamical metric coupled to the SM \cite{1,bbkn}.
In principle, all single-trace operators of the hidden theory  give rise to emergent low-energy fields, coupled to the SM model.
For scalar operators, this is detailed in sections \ref{CoupledSectors} and \ref{axion}. However, most of them acquire effective masses, that are of the order of the messenger mass scale $M$. Therefore, they are irrelevant for low-energy physics.
This is analyzed in detail in section  \ref{axion}.
There are only three classes of operators that are protected by symmetries, and therefore  have  generically low masses.

\begin{itemize}

\item The total conserved stress tensor. It gives rise to emergent gravity, \cite{1,bbkn}.

\item  Instanton densities. They are protected by their topological invariance. They give rise to emergent axions.

\item Conserved global currents. They  gives rise to emergent gauge fields, \cite{1,abck}.

\end{itemize}

There are however, less common symmetries that have not been considered in \cite{1}. An example is  higher form symmetries. The only continuous example we know of in $D=4$ is an antisymmetric tensor conserved current, which can always be mapped to a free U(1) gauge field. Such a current, if generically coupled, will be broken and therefore the associated particle will acquire a mass.
All other examples we know of,  couple to non-local discrete symmetries.

{In the rest of the paper we pursue the topic of emergent axions.
We assume a setup where there is a four-dimensional hidden theory which is coupled to the visible theory (the SM) via a set of bifundamental messenger fields with typical masses of order $M$.
In most of the paper we assume $M$ to be much bigger than any SM scale, or hidden field theory scale\footnote{We will entertain the opposite inequality in the end of section \ref{strong}.}.
Our main goal is a hidden theory with a large N gauge group, and this is the reason why we sometimes call it QFT$_N$. However, we shall assume the large N limit only when necessary.}

\subsection{Results}

 We would like to compute the direct couplings of the operator $A\equiv Tr[F\wedge F]$ in the ``hidden" large-N QFT to the associated operators of the SM.

We expect that
\begin{itemize}

\item It will couple to all analogous operators of the SM $z_I=Tr[F_I\wedge F_I]$ where $I=1,2,3$ labels one of the simple factors of the SM gauge group. It might also couple to other CP-odd scalar operators (like $\bar\psi \psi H-cc$).

\item It will not couple to other CP-even operators\footnote{This assumes, as we do here, that the messenger sector does not break CP.}.

\end{itemize}

Our first goal will be to understand how these couplings are generated.

To do this we must write the couplings of the messenger fields to the SM as well as to the large-N QFT that we denote by QFT$_N$.

We denote  the messenger fields by $\phi_{ai}$, where the index $a=1,2,\cdots, N$ is the color index of the hidden QFT$_N$ while $i$ is a gauge index of a gauge group of the SM. $\phi_{ai}$ can be a vector boson, a fermion or a scalar. Note that in order to be able to write UV-complete messenger couplings to the
SM,  all the SM fields should be bi-fundamentals\footnote{In a generalised sense, encompassing adjoints and rank-two (anti)symmetric tensors.}, $\chi_{ij}$,  with respect to the gauge groups of the SM. Again,  here $\chi_{ij}$ can be a scalar, a fermion or a vector boson. Note that there are many ways to write the SM fields as bifundamentals, but all these ways have been carefully classified in \cite{pa}, and they involve the inclusion of at least one more and typically several  extra U(1)s. These U(1) are generically anomalous. Finally, we label the QFT$_N$ fields as $A_{ab}$ as they are in the adjoint (or bifundamentals) of the large-N gauge group.
There is a set of anomaly cancellation conditions that the messenger sector should satisfy that are presented in appendix \ref{anomalies}.

The couplings of messengers to the SM fields and to the QFT$_N$ fields can be schematically written as
\be
S_{int}=\lambda\int d^4 x ~\left[\phi^*_{ai}\phi_{aj}\chi^{ij}+\phi^*_{ai}\phi_{bi}A^{ab}\right]
\label{0}\ee
where we have suppressed possible derivatives and other space-time indices.
The notation is sketchy and the spin of the various fields is not indicated.
The relevant coupling constants are the appropriate gauge couplings. For now we collectively denote them by $\lambda$.
There may be also quartic couplings in (\ref{0}) but they do not affect our analysis.

Our results can be summarized as follows.
\begin{itemize}
\item Generic couplings in the messenger sector generate interactions of hidden and observable instanton densities. It is expected in general and will be shown by an explicit one-loop calculation in a  typical example with messengers being fermions.

\item There will be many other double (or higher) trace couplings generated by the messenger sector. With the exception of conserved operators discussed earlier, they will lead to interactions that are highly suppressed in the IR ($E\ll M$). The reason is that the corresponding emergent fields will acquire masses of order $M$.

\item We analyse the induced couplings in the observable sector of general scalar-scalar interactions between the hidden and the observable theory in section \ref{CoupledSectors} by assuming a semiclassical (quadratic) approximation. This is valid if both theories belong to one of the following two options: a large N theory or a perturbative theory. The effect of generic higher interactions in this  picture will
be analyzed in section \ref{Interactions} and will be shown that it does not change the leading picture.

\item The inter-theory couplings will be reformulated in section \ref{axion} in terms of scalar fields that represent the associated composite operators of the hidden theory. Their quadratic action will be determined in terms of hidden and observable two-point functions.

\item It will be shown that, for generic scalar operators, the associated composite particles  have ${\cal O}(M)$ masses and are therefore irrelevant in the low-energy effective theory.

\item It will also be shown that for scalar operators that are instanton densities, the effective action for their composites has a different parametric dependence. The hidden instanton density generates an emergent axion coupled to the SM.
    The characteristic decay constant of the emergent axion is, generically,
    \be
    f_a\sim m_{h}\left({M\over m_{ h}}\right)^{4}\gg m_{h}
    \ee
    where $m_h$ is the characteristic scale of the hidden theory.
    The corrections to this scale from SM quantum effects are suppressed.

\item The mass $m_a$ of the emergent axion has two contributions. The first is due to SM quantum effects and is $\sim \Lambda_{QCD}^2/f_a$ as with standard axions.
    However, unlike fundamental axions, the emergent axion mass has also a contribution from the hidden theory that is dominant and of order $m_h$. The reason is that in the theoretical setup we analyze, the associated PQ symmetry is broken by non-perturbative effects both in the hidden and in the visible theory.

\item    In the case where the hidden theory is a CFT, the emergent axion mass and decay constant arises entirely from SM corrections, and their parametric dependence is different
 \be
f_a \sim {\Lambda_{QCD}^3\over M^2}\sp m_a \sim \Lambda_{QCD}
\ee
Therefore the mass is comparable to that of hadrons and moreover it is no longer ``invisible". For energies $\Lambda_{QCD}\gg E\gg f_a$ it is strongly coupled and it is interesting to derive the resulting interaction it mediates\footnote{An analogue of the Vainshtein mechanism, \cite{vain},  may be at play here.}.

\item There are cases, when the characteristic scale of the hidden theory is sufficiently low that the axion has a non-local kinetic term. This is analyzed in section \ref{reg}. Such axions are still interesting and well-defined, however the analysis of the experimental viability is different, and standard experimental constraints do not directly apply.

\item When the hidden theory is a holographic theory then the description that we developed needs a further amendment, so that the Lagrangian description is closer to the dynamics.
    If the instanton density in the hidden theory has a gapped and discrete spectrum as in QCD\footnote{The same description of course works for the case of continuous gapless spectrum, but the discrete spectrum picture is more intuitive.}  then a better ``resolution" of the interaction is to introduce an axion field for each pole of the two-point correlator of the instanton density. This amounts to the presence of an infinite number of four-dimensional axions and as usual the dynamics organizes them into a single five dimensional axion field in the emergent fifth holographic dimension.

The appropriate picture then for the  hidden theory is as a bulk five-dimensional holographic theory that interacts with the SM represented as a brane immersed in the five dimensional space-time at an appropriate radial direction corresponding to the messenger mass cutoff $M$.

This picture is developed in section \ref{Hologaxion}. It resembles the DGP setup, \cite{DGP}, but now the field at stake is an axion. Moreover, the hierarchy between bulk and brane kinetic terms that marred the DGP setup now is naturally explained from the dual QFT picture.

An analysis of the effective interaction mediated by the axion on the brane can be done along the same lines as for the DGP case, \cite{DGP,irs,CKN}.
We find that the interaction at short and long distances is the one of a four-dimensional massive scalar. Depending on some parameters, at intermediate distances there may be a phase where the axion interaction is five-dimensional. In such a regime all emergent axion resonances contribute equally to the relevant interactions.

The obtained parameters in this case are
\be
f^2_{a}=M^2+2(M_P\ell)^3{\bar d_2\over \bar d_0^2}m_h^2\sp m_{a}^2=
{\Lambda^4+2{(M_P\ell)^3\over \bar d_0}m_h^4\over f_{a}^2}\;.
 \label{bb17}\ee
 where $\Lambda,M$ are (brane) parameters defined in (\ref{b5}), $M_P,\ell, m_h$ are the bulk parameters. $\bar d_n$ are dimensionless numbers, typically of order one.

\item An analysis of experimental constraints, performed in section \ref{PhenoConsiderations} and where it applies, indicates that composite axions that are instanton densities, in the paradigm proposed in \cite{1}, cannot be QCD-axions, except in the holographic case as anticipated.
    They can be on the other hand inflatons even if the hidden theory is not holographic. However there exist special regimes even in the non-holographic cases, for which a more thorough analysis needs to be performed due to the substantially different kinetic terms that such emergent axions have.


\end{itemize}

There are many directions that remain open and are related to the ideas presented in this paper.
The phenomenology and potential model building of emergent axions is unexplored. In reference \cite{1} it was argued that one or more massive anomalous U(1)'s may form part of the SM as in the case of orientifolds, \cite{pa,class,AKT,KA,rev}. Their low energy effective action was analyzed  in \cite{CIK}. The interplay of the emergent axions with the anomalous U(1) gauge bosons is interesting to explore and the string theory knowledge is a good guide in this direction \cite{a1}-\cite{a3}.

The cosmology is also novel as in the holographic case the description is quite different from the standard one. The role of the emergent axion as an inflaton may be investigated along the lines of \cite{Def}.
The role of such  axions as (non-local) dark matter is also interesting to analyze.

The emergent axion may also play the role of a relaxion, \cite{relax}. Indeed the bulk axion  can combine with other relevant perturbations of the bulk theory, along the lines of \cite{CKN} to provide a different mechanism for relating the hierarchy problem associated with the EW scale and the self-tuning of the cosmological constant \cite{hknw}.

Other emergent interactions along the lines of \cite{1} will be discussed elsewhere, \cite{bbkn,abck}.

\section{The effective cross-couplings}

We start by considering the  messenger interactions between the hidden QFT$_N$ and the visible theory (which for us will be the SM). We perform a simple calculation to assess the emergent double trace couplings that will be our focus in this paper.
We consider in particular the effective terms in the action which are generated by messengers going around a loop.

If we only have QFT$_{N}$ external states, such corrections will affect the action of QFT$_N$.  At large N, this correction is suppressed by 1/N compared to the leading corrections due to the hidden theory itself, as the messengers transform in the fundamental of SU(N).
If we only include SM external fields, then we obtain corrections to the SM couplings that are ${\cal O}(N)$.\footnote{It is interesting to note that at large N, such corrections can change substantially the SM couplings at the cutoff scale $M$ (defined as the messenger decoupling scale).}

We are interested in corrections that involve external fields from both sides as these will generate interactions in the IR between  QFT$_N$ and the SM.
By gauge invariance, the minimal numbers of fields we need on each side is two, so the leading correction is generated by box diagrams.

Our setup is perturbative, but also indicative of the general case, that is not necessarily perturbative.

\subsection{The exceptional couplings}

Before we discuss the generic case, we start from the exceptions.
They involve U(1) gauge fields that can be made gauge invariant with a single field strength , namely $F_{\m\n}$.
There are good reasons to believe that QFT$_N$ does not have $U(1)$ gauge fields  as they decouple in the holographic limit.
However, the SM has the hypercharge, as well as possible anomalous U(1)'s.
Couplings of $F^Y_{\m\n}$ to appropriate $QFT_{N}$ gauge invariant operators are special as $Tr[Q_Y]=0$. They will be discussed in a companion paper \cite{abck}.

There can be couplings also to the SM anomalous U(1)'s. We shall not consider them here.

\subsection{The generic couplings}\label{genericCoupling}

We consider the generic couplings induced between QFT$_N$ and SM.
They involve single-trace, gauge invariant operators of QFT$_N$ of any spin coupled to similar gauge invariant operators of the SM.

We focus on operators involving two non-abelian gauge fields in QFT$_N$ and two gauge fields on the SM side. The reason is that such operators include the very special case of the instanton densities. They are special as they are protected by perturbative shift symmetries and they are expected to  induce effective emergent axion couplings on the SM side \cite{1}.
This class of calculations is in the ``light by light scattering" class and various aspects have been addressed in the literature, \cite{Kar}-\cite{RD}.

We insert therefore two gauge fields from QFT$_{N}$ and two from the SM.
The terms we are interested in will come from fermion messengers going around the loop. Similar terms are expected to appear also from bosonic messengers. We shall not study bosonic messenger contributions here, as they are expected to be similar.

The effective Lagrangian obtained describes double trace interactions between the hidden theory QFT$_N$ and the SM.
The relevant operators that participate are
scalars and four index tensors
\be
G^a_{\m\n}G^{a,\m\n}\sp
G^a_{\m\n}\widetilde G^{a,\m\n}\sp
G^a_{\m\n}G^{a}_{\r\s}\sp
G^a_{\m\n}\widetilde G^{a}_{\r\s}
\ee
in the hidden theory and the corresponding ones in the SM,
\bea
&&F^i_{\m\n}F^{i,\m\n}\sp
F^i_{\m\n}\widetilde F^{i,\m\n}
\sp F^i_{\m\n}F^{i}_{\r\s}\sp
F^i_{\m\n}\widetilde F^{i}_{\r\s}
\eea
The final one-loop interaction Lagrangian can be written as
\bea
S_\textup{eff}&=&
-\frac{g_\textup{SM}^2 g_\textup{QFT}^2}{90(4\pi)^2 M^4}\int d^4x\Big[ G^a_{\m\n}G^{a,\m\n} F^i_{\r\s}F^{i,\r\s} +2 G^a_{\m\n}G^{a}_{\r\s} F^{i,\m\n}F^{i,\r\s} + \nn\\
&& ~~~~~~~~~~~~~~~~~~~~~~~~~~~~+\frac{7}{4} G^a_{\m\n}\widetilde G^{a,\m\n} F^i_{\r\s}\widetilde F^{i,\r\s} +\frac{7}{2} G^a_{\m\n}\widetilde G^{a}_{\r\s} F^{i,\m\n}\widetilde F^{i,\r\s} \Big]
\label{final}\eea

The large N estimate of this calculation agrees with the generic large N estimates made in appendix A of \cite{1} provided there is a redefinition of the normalization of the operators so that the two normalization agree.
The detailed computation is shown in appendix \ref{boxAppendix}.

At large N, and strong coupling, the tensor fields are expected to acquire large anomalous dimensions\footnote{In the holographic dual, they are typically represented by stringy states.} and are not therefore expected to be important at low energy. The scalar coupling, as we show in the next section will also not be very relevant at low energy. It is only the coupling of the instanton densities that will turn out to be important and give rise to an emergent axion.

\section{The general scalar cross interaction and its emergent resolution}\label{CoupledSectors}

We would now like to reinterpret the interaction between two scalar operators in the two theories and in particular the instanton densities, in an effective description. To do this,  we would like to address the more general problem of a scalar-scalar interaction between two theories and its IR ``resolution".

We consider two theories $T_1$ and $T_2$  coupled via an interaction of the form
\be
S_{12}=\l\int d^4x ~O_1(x) O_2(x)
\label{a1}\ee
where $O_1$ is an operator of dimension $\Delta_1$ belonging to $T_1$ and $O_2$ is an operator of dimension $\Delta_2$ belonging to $T_2$.
This may be an UV coupling, defining an interacting pair of theories in the ultimate UV.
It may be also the definition of the coupled theory at a finite energy cutoff. For instance this is the case at the scale of the messenger mass $M$.
In that case
\be
\l=\l_0 M^{4-\Delta_1- \Delta_2},
 \label{c1}\ee
 with $\l_0$ dimensionless.
We assume that the hidden theory $T_1$ is a theory at large N. We shall explicitly show the dependence on $N$ in the sequel. In particular, the coupling $\l_0\sim {\cal O}(1)$ when $O_1$ is a standard normalized single trace operator so that all its correlators are of order ${\cal O}(N^2)$.

We assume that the two theories are such that higher than three-point (connected) functions are suppressed  compared to two-point functions.
Examples of such theories are near-free theories (interactions suppressed by small couplings) and large N-theories (interactions of gauge-invariant operators suppressed by 1/$N$).

Consider now the generating functional for the correlators of $O_1,O_2$
 \be
Z(J_1,J_2)=\langle 0|e^{i S_{12}+i \int d^4x ~(J_1(x) O_1(x)+J_2 (x) O_2 (x))}|0\rangle\sp  e^{i W(J_1,J_2)}\equiv {Z(J_1,J_2)\over Z(0,0)}
\label{a2} \ee
By performing a Hubbard-Stratonovich transformation we can write
\be
e^{i S_{12}}=N_0\int {\cal D}\zeta_1 {\cal D} \zeta_2 ~e^{\int d^4x~\left(-{i \over \l}\zeta_1(x) \zeta_2(x) - i\zeta_1 O_1 - i \zeta_2  O_2\right)}
\label{a3}
\ee
that allows us to express the complete generating functional as follows
\be
Z(J_1,J_2)=\langle 0|e^{i S_{12}+ i \int d^4x ~(J_1(x) O_1(x)+J_2 (x) O_2 (x))}|0\rangle =
\label{b1}\ee
$$
= N_0\int {\cal D}\zeta_1 {\cal D} \zeta_2~e^{\int d^4x~\left(-{i\over \l}\zeta_1(x) \zeta_2(x)\right)}Z_1(J_1 -  \zeta_1) Z_2(J_2 -  \zeta_2)
$$
where $Z_1, Z_2$ are the Schwinger functionals of the respective uncoupled theories
\be
Z_1(J_1)=\langle 0|e^{i \int d^4x ~ J_1(x) O_1(x)}|0\rangle_1 \sp Z_2(J_2)=\langle 0|e^{i \int d^4x ~J_2 (x) O_2 (x)}|0\rangle_2 \sp
\ee
We henceforth work at the quadratic order\footnote{We have assumed that operators have zero vevs. Corrections to the semi-classical approximation will be discussed in  section~\ref{Interactions}.} in which
\be
Z_1(J_1)=e^{{i \over 2}\int d^4x d^4 x'~J_1(x)J_1(x')G_{11}(x-x')}=e^{{i \over 2}\int {d^4p\over (2\pi)^4}J_1(p)J_1(-p)G_{11}(p)}
\label{b2}\ee
with $G_{11}(x-x') = \langle O_1(x) O_1(x') \rangle_1$ is the translationally invariant, unperturbed two-point correlation function of $O_1$ in theory $T_1$ and with a similar expression defining $Z_2$.
Using (\ref{b2}) and performing the integral over $\zeta_1, \zeta_2$ in (\ref{b1}) explicitly, we obtain the quadratic order generating functional (expressed in momentum space) as
\be
W(J_1,J_2)=
{i\over 2}\int {d^4 p\over (2\pi)^4}\left[\left(\begin{matrix} J_1(p),  & J_2(p)\end{matrix}\right)\left(\begin{matrix}{1\over G_{11}(p)}& -{\l}\\
-{\l}& {1\over G_{22}(p)}\end{matrix}\right)^{-1}\left(\begin{matrix} J_1(-p)\\ J_2(-p)\end{matrix}\right)\right]
\label{b3}\ee
Equivalently one can use the saddle point equations (\ref{b1}) (since they are exact for Gaussian integrals)
\bea
 \zeta_1(x) + \l\int d^4y~  G_{22}(x-y)(\zeta_2(y) - J_2(y)) &=& 0\, , \nn \\
 \zeta_2(x) + \l\int d^4y~  G_{11}(x-y)(\zeta_1(y) - J_1(y)) &=& 0
\label{b4}
\eea
and substitute them back to obtain (\ref{b3}).
More explicitly,  one can rewrite the matrix appearing in (\ref{b3}) as
\be
\left(\begin{matrix}{1\over G_{11}(p)}& -{\l}\\
-{\l}& {1\over G_{22}(p)}\end{matrix}\right)^{-1}={1\over 1-{\l^2} G_{11}(p) G_{22}(p)}\left(\begin{matrix}G_{11}(p)& {\l}G_{11}(p) G_{22}(p)\\
{\l}G_{11}(p) G_{22}(p)&  G_{22}(p)\end{matrix}\right)
\label{a9}\ee
The scaling dimension of $G_{11}(p)$ is $2(\Delta_1-2)$ and the one of $G_{22}(p)$ is $2(\Delta_2-2)$.
We notice that the interaction between the two theories modifies the non-interacting correlators and create cross-correlations between the two sectors. As an example, the new correlator for $O_2$ in momentum space is
\be
i\langle O_2(p) O_2(-p) \rangle={G_{22}(p)\over  1-{\l^2} G_{11}(p) G_{22}(p)}= G_{22}(p)+{\l^2}{{G_{11}(p)}{ G_{22}^2(p)}\over  1-{\l^2} {G_{11}(p)}{ G_{22}(p)}}
\label{a10}\ee
In particular, in the regime where $G_{11} \to 0$, it is given by the initial correlator $G_{22}$ while in the regime where $G_{11} \to \infty$, it is given by $-1/ \l^2 G_{11}$.

We now consider the IR expansion of the correlators. It depends crucially on the detailed physics of each theory.
In a theory with a single scale, which is also its mass gap $m$ (like YM), the IR expansion in $p\ll m$ reads
\be
i\langle O_2(p) O_2(-p) \rangle_2 =  b_0 +  b_2 p^2+  b_4p^4 +\cdots\sp b_n\sim m^{2(\Delta_2-2)-n}
\label{a11}\ee
If $\Delta_2$ is an integer, then starting with the term $p^{2(\Delta_2-2)}$ logs of momentum appear in the expansion.

On the other hand, for $p\gg m$ the UV expansion of the (renormalized) correlator in momentum space is
\be
i\langle O_2(p) O_2(-p) \rangle_2 =p^{2(\Delta_2-2)}\left[\log{p^2\over m^2}\left(a_0+a_2{m^2\over p^2}+{\cal O}\left({m^4\over p^4}\right)\right)+\right.
\label{a11a}\ee
$$\left.
+c_0+c_2{m^2\over p^2}+{\cal O}\left({m^4\over p^4}\right)\right]
$$
where the coefficients $a_i,c_i$ are dimensionless. The expansion (\ref{a11a}) is valid whether $\Delta_2$ is integer or not.
This expansion reflects the fact that at short distances $p\gg m$ the correlator in configuration space asymptotes to the CFT value, proportional to $|x|^{-2\Delta_2}$.

If the theory has a UV scale $\Lambda$, but also other smaller IR scales like $m\ll \Lambda$ then for generic scalar operators the larger scale dominates the coefficients in the expansion (\ref{a11}),
\be
b_n\sim \Lambda^{2(\Delta_2-2)-n}\left[1+{\cal O}\left({m^2\over \Lambda^2}\right)\right]
\ee
There is however a scalar operator in the gauge theory that is special and for which this scaling is not valid. This is the (CP-odd) instanton density operator that is the focus of the present paper. It is well known from studies in QFT, \cite{panago} and holography, \cite{ihqcd} that the correlators of the instanton density are UV insensitive.
The reason is that the $\theta$ angle in QCD is not renormalized, as shown rigorously on the lattice (see \cite{ihqcd}-\cite{DiVecchia:1981aev} for a detailed discussion).
This is also true in holography, whereby the bulk axion field dual to the instanton density does not have a potential and the procedure of holographic renormalization allows to derive its correlation functions \cite{Bianchi:2001kw, Bianchi:2001de}.  However, in holographic QCD there is a non-trivial (and non-perturbative)  $\beta$-function for $\theta$ driven by the vev of the instanton density on the (non-trivial) YM vacuum (encoded in the topological susceptibility), \cite{ihqcd}.
Notwithstanding this, all $\theta$-dependent contributions to the vacuum energy are cutoff independent.

There is a quick check of this fact when we consider the contribution of heavy quarks (or scalar quarks) with masses $M\gg \Lambda_{YM}$ to the two-point function of the instanton density. Assuming perturbation theory to hold, the calculation is a variant of the one-loop calculation presented in appendix \ref{boxAppendix}. The results indicate that the contributions vanish in the limit  $M\gg \Lambda_{YM}$ as $M^{-4}$. This was also calculated from first principles in \cite{NSVZ}.

Therefore, even though there is a non-trivial UV structure in the gauge theory, the two-point function scales as in (\ref{a11}) where $m$ is the characteristic IR scale of the gauge theory, and UV scales (like the messenger mass) do not appear in the correlator.
{\em This is an important feature that distinguishes this operator from all other scalar operators and has important IR consequences as we shall see further on.}

We also similarly parametrize in the IR
\be
i\langle O_1(p) O_1(-p) \rangle_1  \simeq ~ a_0 +a_2p^2+a_4p^4 +\cdots
\label{a14}\ee
an expansion valid when $p\ll m_1$. Like the previous discussion, if $O_1$ is the instanton density, the size of the coefficients $a_n$ is controlled mainly  by $m_1$.

If there is a non-trivial CFT in the IR, then there are also non-analytic contributions. For example, for an operator with IR dimension $\Delta_{IR}$ we have instead

\be
i\langle O_1(p) O_1(-p) \rangle_1  \simeq  a_0 +a_2p^2+a_4p^4 +\cdots+ p^{2(\Delta_{IR}-2)}\log p^2\left(a^1_0+a^1_2p^2+\cdots\right)+\cdots
\label{a14a}\ee
For a theory like QCD, with mass gap, such non-local contributions are absent in the IR.

The dependence of the coefficients $a_n$ on the various scales of the theory follow our discussion above for $\langle O_2(p) O_2(-p) \rangle_1$.

The case that one of the two theories is a CFT, must be discussed separately, although the main qualitative behavior can be inferred from what was mentioned so far.
In an exact CFT, the two-point function of an appropriately normalized  scalar operator in configuration space is given by
\be
\langle O(x)O(0)\rangle={1\over |x|^{2\Delta}}
\ee
The Fourier transform of this is ill-defined for $\Delta>2$ in four dimensions and it is well known that a regularisation is needed.
Using a short-distance cutoff $\Lambda$ one obtains
\be
\langle O(p)O(-p)\rangle\sim p^{2(\Delta-2)}\left({p\over \Lambda}\right)^{2-\Delta}K_{\Delta-2}\left({p\over \Lambda}\right)
\ee
valid for $p\ll \Lambda$. For $\Delta>2$ it exhibits a mild (logarithmic) UV divergence associated with the log in the K-function.
This structure  can be obtained by taking the limit $m\to \infty$ of (\ref{a11a}).

We will comment here on the $N$-dependence of the various functions and parameters in the case the hidden theory $T_1$ is a large N theory.
As was shown in \cite{1}, the interaction (\ref{a1}) has $\l\sim O(1)$ when the operator $O_1$ is normalized so that
\be
\langle O \rangle \sim N\sp \langle O O\rangle \sim {\cal O}(1)\sp   \langle OOO\rangle \sim {\cal O}(N^{-1})\sp ...
\ee
Therefore $\l$ and $G_{11}$ are ${\cal O}(1)$.

\section{Integrating in a new (pseudo)-scalar\label{axion}}

We would like now to interpret the presence of the interaction (\ref{a1}) from the point of view of theory $T_2$ (that from now on we will call $T_{_{SM}}$ to indicate that it is the observable theory) as due to a novel dynamical scalar, coupled linearly to the operator $O_{2}\equiv O_{_{SM}}$.
In a sense, we consider $T_1$ as the ``hidden" theory and from now on we will denote it as $T_{h}$. The new coupling is a channel of communication between the hidden and the visible theory induced by the interaction (\ref{a1}) in the UV.

Therefore we imagine that we probe theory $T_{_{SM}}$ and we can perform experiments involving only the operator $O_{_{SM}}$ of $T_{_{SM}}$\footnote{In the general case, the interaction between the two theories involves multiple operators. The generalisation of our calculation to any number of operators for the theory $T_{_{SM}}$ is straightforward.}. We wish to ask to which extent we can represent the effects of $T_{h}$ and its interaction to $T_{_{SM}}$ as coming from an ``emergent" dynamical field coupled linearly to $O_{_{SM}}$.
We henceforth call the theory $T_h$ as the ``hidden sector" while the theory $T_{_{SM}}$ will represent for us the visible sector.

We consider a new scalar field $\chi$ coupled to the operator $O_{_{SM}}$ as follows
\be
S_{eff}=\int d^4x\left[{1\over 2}\chi K \chi+g\chi O_{_{SM}} \right]+ S_{_{SM}}=\int {d^4p\over (2\pi)^4}\chi(p)K(p)\chi(-p)+g\chi(p)O_{_{SM}}(-p)+S_{_{SM}}
\label{a16}\ee
where $K$ is an  operator that we want to determine using consistency with the results of section~\ref{CoupledSectors}.
We have inserted a dimensionfull coupling $g$ in the interaction of $\chi$ with $O_{_{SM}}$ so that the scalar $\chi$ is dimensionless.
Because of this, the mass dimension of $g$ is   $4-\Delta_{_{SM}}$ and the operator $K$ has scaling dimension 4 in mass.

In order to determine the correct form of $K$, we now compute the $O_{_{SM}}$ correlator by integrating out the scalar field. To do this properly we introduce again a source for $O_{_{SM}}$
\be
Z(\theta_2)=\int {\cal D}\chi\langle  e^{i S_{eff} + i \int d^4 x ~\theta_{_{SM}}(x) O_{_{SM}}(x)}\rangle
\sp  e^{i W(\theta_{_{SM}})}\equiv {Z(\theta_{_{SM}})\over Z(0)}
\label{a18}\ee
From the definitions we have
\be
\langle  e^{i S_{eff}+i \int d^4 x ~\theta_{_{SM}}(x) O_{_{SM}}(x)}\rangle=e^{i \int d^4x{1\over 2}\chi K \chi}~N~\times
\label{a19}\ee
$$\times
e^{{i \over 2}\int {d^4p\over (2\pi)^4}(g\chi(p)+\theta_{_{SM}}(p))G_{{_{SM}},{_{SM}}}(p)(g\chi(-p)+\theta_{_{SM}}(-p))+{\cal O}((g\chi+\theta_{_{SM}})^3)}
$$
that allows us to compute at quadratic order
\be
Z(\theta_{_{SM}})=N'e^{{i\over 2}\int {d^4p\over (2\pi)^4}\theta_{_{SM}}(p)\theta_{_{SM}}(-p){G_{{_{SM}},{_{SM}}}(p)K(p)\over K(p)+g^2G_{{_{SM}},{_{SM}}}(p)}}
\label{a20}\ee
and
\be
W(\theta_{_{SM}})={1\over 2}\int {d^4p\over (2\pi)^4}\theta_{_{SM}}(p)\theta_{_{SM}}(-p){G_{{_{SM}},{_{SM}}}(p)K(p)\over K(p)+g^2G_{{_{SM}},{_{SM}}}(p)}
\label{a21}\ee
Differentiating twice with respect to $\theta_{_{SM}}$ we obtain
the corrected two-point function for $O_{_{SM}}$
\be
i \langle O_{_{SM}} O_{_{SM}} \rangle={G_{{_{SM}},{_{SM}}}(p)K(p)\over K(p)+g^2G_{{_{SM}},{_{SM}}}(p)}=G_{{_{SM}},{_{SM}}}(p)-g^2{G_{{_{SM}},{_{SM}}}(p)^2\over K(p)}+\cdots
\label{a17} \ee
If we wish to reproduce the two-point function obtained in the previous section, in (\ref{a10}), we must match it to (\ref{a17}).
We find
\be
K(p)=-{g^2\over \l^2G_{h,h}}
\label{a22}\ee
This expresses the consistency relation between the two descriptions.

\subsection{The strong CP problem revisited \label{strong}}

{We should first address to what extend the novel effective pseudoscalar $\chi$ should be called an ``axion" and to what extend it is an ALP (axion-like particle) or the QCD axion.
 To assess this we must remember the qualifying properties of an  axion-like particle: a linear coupling to the QCD instanton density, and an approximate U(1) symmetry, broken by non-perturbative effects guarantee that we have an ALP.  The linear coupling to the QCD instanton density is expected from the general arguments we presented earlier, as well as the calculations we presented in appendix \ref{boxAppendix}. The associated U(1) theory is true in perturbation theory and it is the perturbative symmetry  shifting the $\theta$ angle. Therefore we definitely have an ALP.

To see to what extend we have a QCD axion, we need one more property, namely that the minimum of the axion potential is at zero. This is guaranteed for the potential generated by QCD or any real gauge theory by general considerations \cite{Witten:1979vv} and the Vafa-Witten theorem, \cite{VW}.
However, the presence of two relevant theories, introduces an extra context.

\begin{itemize}

\item Consider, as a simple  example two YM theories with scales $\Lambda_h,\Lambda_{_{SM}}\ll M$ coupled at the messenger scale by a coupling of the form derived in appendix \ref{boxAppendix}
\be
S_{CP-odd}=\theta_h\int O_h+\theta_{_{SM}}\int O_{_{SM}}+\int {O_hO_{_{SM}}\over M^4}
\label{a22d}\ee
where $O_{h,{_{SM}}}$ are the two instanton densities.
Resolving the interaction using the Hubbard-Stratonovich trick we obtain
\be
S_{CP-odd}=(\theta_h+s_1)\int O_h+(\theta_{_{SM}}+s_2)\int O_{_{SM}}-M^4s_1s_2
\label{a22e}\ee
where $s_{1,2}$ are two auxiliary dynamical variables.
The effective potential for the theta-angles of the combined theory follows from  (\ref{a22e})
\be
V_{eff}=V_h(\Lambda_h,\theta_h+s_1)+V_{_{SM}}(\Lambda_{_{SM}},\theta_{_{SM}}+s_2)-M^4s_1s_2
\label{a22f}\ee

The structure of the potentials $V_{h,{_{SM}}}$ depends on whether the YM theories have light quarks or not. However in all cases, the VW theorem states that the global minimum of the faction $V_i(\Lambda_i,x)$ is at $x=0$.

The strong CP-problem is absent if the relevant vev of the instanton density
\be
\langle O_i\rangle=V'_i(\Lambda_i,\theta_i+s_i)\Big|_{extremum}
 \ee
 vanishes (or is sufficiently small in appropriate units) at the extremum.
In pure YM theory
\be
V(\Lambda,\theta)=\Lambda^4f(\theta)\sp f(\theta+2\pi)=f(\theta)
\label{a22h}\ee
where $f(\theta)$ is a periodic function with an absolute minimum at $\theta=0$, \cite{Witten:1979vv}. At large $N_c$ the function becomes quadratic and discontinuous. In the presence of light quarks of mass $m_q\ll \Lambda$
\be
V(\Lambda,\theta)\simeq \Lambda^3m_q\cos(\theta)+{\cal O}(m_q^2)
\ee
In all of the above the vev of the instanton density is proportional to $f'(\theta)$ and vanishes only at $\theta=0$.

The $s_{1,2}$ equations emerging from (\ref{a22f}) are\footnote{For $M\gg \Lambda_{1,2}$ they are a good approximation to the exact answer.}
\be
{V'_h(\Lambda_h,\theta_h+s_1)\over M^4}=s_2\sp {V'_{_{SM}}(\Lambda_{_{SM}},\theta_{_{SM}}+s_2)\over M^4}=s_1
\label{a22g}\ee
If $M\gg \Lambda_{h,{_{SM}}}$ then the solutions to (\ref{a22g}) are
\be
s_1={V'_{_{SM}}(\Lambda_{_{SM}},\theta_2)\over M^4}+{\cal O}\left({\Lambda_{h,{_{SM}}}^4\over M^4}\right)\sp s_2={V'_h(\Lambda_h,\theta_h)\over M^4}+{\cal O}\left({\Lambda_{h,{_{SM}}}^4\over M^4}\right)
\ee
and therefore
\be
\langle O_i\rangle=V'_i(\Lambda_i,\theta_i) +{\cal O}\left({\Lambda_{h,{_{SM}}}^4\over M^4}\right)
 \ee
We conclude that in this case the emergent axion does not solve the strong CP-problem.

 \item  $\Lambda_h\gg M\gg \Lambda_{_{SM}}$. If $M^2\gg \Lambda_h\Lambda_{_{SM}}$ then there is a unique solution to the saddle point equations
      while in the opposite case there are ${\cal O}\left({M^4\over \Lambda_{_{SM}}^4}\right)$ solutions.
However in all of them $s_{h}\leq {\cal O}\left({M^4\over \Lambda_{_{SM}}^4}\right)$ and it turns out that the strong CP-problem is not resolved.
The same applies to the opposite hierarchy $\Lambda_h\ll M\ll \Lambda_{_{SM}}$.

\item $\Lambda_h,\Lambda_{_{SM}}\gg M$. In this case the saddle-point equations (\ref{a22g}) have many solutions.
     The ones relevant for resolving the strong CP problem can be written as
   \be
    s_{1}\simeq -\theta_h-\theta_{_{SM}}{M^4\over V_h''(0)}+{\cal O}\left({M^8\over V_h''(0){V_{_{SM}}}''(0) }\right)\;,
\ee

 \be
    s_{2}\simeq -\theta_{_{SM}}-\theta_{h}{M^4\over V_{_{SM}}''(0)}+{\cal O}\left({M^8\over V_h''(0){V_{_{SM}}}''(0) }\right)\;,
\ee
and instanton vev that characterizes the resolution of the strong CP problem is given by
\be
{\langle O_{_{SM}}\rangle \over \Lambda_{_{SM}}^4}\sim {M^4\over V_{_{SM}}''(0)}\ll 1
\ee

For a gauge theory with light quarks $f(\theta)=\cos\theta$ in (\ref{a22h}) and the formulae above become

    \be
    s_{1}\simeq -\theta_h+(-1)^{n_1}(\pi n_2-\theta_{_{SM}}){M^4\over \Lambda_h^4}+{\cal O}\left({M^8\over \Lambda^4_h\Lambda^4_{_{SM}}}\right)\;,
\ee
\be
s_{2}\simeq -\theta_{_{SM}}+(-1)^{n_2}(\pi n_1-\theta_{h}){M^4\over \Lambda_{_{SM}}^4}+{\cal O}\left({M^8\over \Lambda^4_h\Lambda^4_{_{SM}}}\right)\;.
\ee
where $n_1,n_2$ integers.

Therefore, only in this last case the composite axion can be  the QCD axion. However, in this case,
strictly speaking our premises are not as assumed in the beginning.
 As the messenger mass is low, the proper description of the axion is as an exotic $\eta'$ and we will review this in section \ref{eta}.

There are other possibilities that those discussed above that are non-generic. They involve the subtleties of another operator that participates in the strong-CP problem, namely the $\eta'$.
Like the instanton density it is protected in perturbation theory, but not non-perturbatively. However in this case there is another parameter, the number of colors, that enters importantly in the story. We will discuss this in more detail in section \ref{eta}.

Finally, there is the possibility, first advocated by Polyakov that the gauge-interactions screen $\theta$.
This has been recently (qualitatively) realized in holographic models of YM, \cite{hknw,ihqcd,data}.
It can be argued, on rather general lines, that the bulk axion field, dual to the $\theta$-angle vanishes in the IR, \cite{hknw,ihqcd,data}.
It is not clear whether this is enough to solve the strong CP Problem, but can be probably combined with other mechanisms in order to increase the ``quality" of the resolution of the strong CP problem.

\end{itemize}
}

\subsection{The IR structure\label{IR}}

To study the physical properties of the field we integrated in, we now consider a generic expansion for the IR behaviour of two-point functions.
As mentioned both theories have a UV scale $M$ that is the messenger scale. Therefore, for generic operators the IR behavior of correlators below the mass gaps of the respective theories  is as follows
\be
iG_{h,h}(p)=a_0+a_2p^2+a_4 p^4+\cdots\sp  iG_{{_{SM}},{_{SM}}}(p)=b_0+b_2p^2+b_4 p^4+\cdots
\ee
Using (\ref{c1}) we obtain
\be
iK(p)={g^2 M^{2(\Delta_h+ \Delta_{_{SM}}-4)}\over N^2\l_0^2}{1\over a_0+a_2p^2+a_4 p^4+\cdots}\simeq {M^{2\Delta_1}\over \l_0^2 a_0}\left[1-{a_2\over a_0}p^2+{a_2^2-a_0a_4\over a_0^2}p^4+\cdots\right]
\label{a23}\ee
where we have chosen the convention
\be
g=M^{4-\Delta_2}
\label{conv}\ee
as the emergent semiclassical axion is ``massless" and  non-local so that the field $\chi$ has mass dimension zero.

Parametrizing then
\be
iK(p)=f_a^2(p^2+{m_a}^2)+{\cal O}(p^4)
\label{a24}\ee
we obtain
\be
m_a ^2\sim {a_0\over a_2}\sp f_a^2\sim {1\over \l_0^2}{a_2\over a_0^2} M^{2\Delta_1}
\label{a25}\ee

For a generic scalar operator $O_1$, as argued before, we have $a_n\sim M^{2(\Delta_1-2)-n}$ and we obtain
\be
m_a^2\sim M^2\sp f_a^2\sim {M^2\over \l_0^2}
\label{a26}\ee

It is clear that the induced interaction is weak but the mass scale of the scalar is a very high scale, the messenger scale.

On the other hand if the operators $O_{h,{_{SM}}}$ are the instanton densities, then as we have mentioned earlier their two-point function is not UV sensitive and in this case
\be
a_{n}=\bar a_n~m_h^{2(\Delta_h-2)-n}\sp b_{n}=\bar b_n~m_{_{SM}}^{2(\Delta_{_{SM}}-2)-n}
\label{a27}\ee
where $m_{h,{_{SM}}}$ are the IR mass scales of the hidden and visible theories $T_{h,{_{SM}}}$ and $\bar a_n,\bar b_n$ are dimensionless and typically ${\cal O}(1)$ coefficients. If the hidden  theory were YM then $m_h$ is $\Lambda_{YM}$.
In this special case we obtain instead
\be
m_a^2={\bar a_0\over \bar a_2} m_h^2\sp f_a^2={\bar a_2\over \bar a_0^2}{m_h^2\over \l_0^2}\left({M\over m_h}\right)^{2\Delta_h}
\label{a28}\ee
If $m_h\ll M$ then this is an emergent weakly-coupled axion-like  field that as we have shown earlier is coupled to the SM instanton densities (which are represented here by $O_{_{SM}}$).

As mentioned above, the case of interest here is when $O_h$ is the ``hidden" instanton density and the $O_h$ is one of the observable (SM) instanton densities. In this case $\Delta_h=\Delta_{_{SM}}=4$ to a high accuracy. We have assumed that $m_h\ll M$. If $m_h$ is also comparable or smaller than SM scales, then this resembles a standard PQ axion as far as the mass is concerned. It has however a compositeness scale that affects its low energy properties.

Equation  (\ref{a28}) gives the dominant contribution to its mass as $m_{h,{_{SM}}}\ll M$.
This is definitely a different situation compared to a fundamental axion field.
{\em Its origin here is not in a continuous global symmetry of a QFT at a higher scale, but an (approximate) ``emergent PQ symmetry" arising from a hidden instanton density.}
Note however that even conventional fundamental axions will have a similar profile when we take into account the expectation that in a theory that contains gravity, all global symmetries expected to be broken by quantum gravitational effects. This in the past has raised the question of the quality of the PQ axion and has been discussed in \cite{quality}.

For the instanton densities, as $\Delta$ is an integer, the low energy structure of the two-point functions is of the form presented in (\ref{a14a}). Therefore, a non-analytic term appears in the emergent axion induced terms a fourth order in derivatives.

There is an interesting limit to discuss, corresponding to  the hidden theory being  a CFT, with $m_h\to 0$. In that case, the only non-trivial scale is the messenger scale that breaks scale invariance in the UV and according to our earlier discussion, in such a case the $\langle O_hO_h\rangle$ correlator\footnote{We assume that there are no spurious contact terms in the correlator and it is defined in agreement with conformal invariance.} starts in the IR as $p^4\log p^2$. The emergent semiclassical axion is ``massless" and  non-local as the inverse propagator starts at ${\cal O}(p^4\log p^2)$.
 This is reminiscent of the Witten-Weinberg theorem, \cite{ww}, although this theorem  applies to emergent massless gravitons and photons\footnote{In the case of emergent gravitons and photons it can be shown that when they are massless, their effective theory is non-local in agreement with the Witten-Weinberg theorem, \cite{bbkn}.}.
Our results suggests that there must be an analogue of this for massless axions.

The SM quantum effects associated to the SM instanton densities provide corrections to the axion action.
In perturbation theory they  provide a renormalization of the kinetic terms but no mass renormalization. The associated one loop diagram is calculated in appendix \ref{apa}.
Non-perturbative QCD effects will provide also a mass correction of order $\Lambda_{QCD}^2/f$ that will be added to the axion mass originating in the hidden theory. The EM instanton density is not expected to contribute to the axion mass.

We can discuss such corrections in general by computing the
 effective mass of $\chi$ which is affected by the mixing with the visible theory  operator $O_{_{SM}}$.
Indeed, a calculation of the two-point function of $\chi$ gives
\be
-i\langle \chi \chi\rangle(p)={-i\over K(p)+g^2 G_{{_{SM}},{_{SM}}}(p)}={\l^2\over g^2}~{1\over {1\over iG_{h,h}}+\l^2 iG_{h,h}}={1\over {1\over \l^2~iG_{h,h}}+iG_{{_{SM}},{_{SM}}}}
\label{a29}\ee
where in the last step we set $g=1$.

When one or both of the operators $O_h$ are generic, the effective mass of $\chi$ is the messenger scale, $M$.
Therefore both $f_a^2$ and $m_a^2$ remain of order ${\cal O}(M^2)$. One can in principle imagine a fine-tuned situation where $m_a\ll M$ but this is not justified at this point.

If both $O_h$ and $O_{_{SM}}$ are UV-protected operators (instanton densities), we can expand\footnote{Note that the behavior of the instanton density two-point function is unusual. In the Euclidean domain, reflection positivity implies that the correlator is negative definite. However, it is known that the topological susceptibility (related to the IR limit of the correlator) is positive. This is due to positive contact terms that exist for this correlator, \cite{panago}.}
\be
iG_{h,h}(p)=m_h^{2\Delta_h-4}\left[\bar a_0-\bar a_2{p^2\over m_h^2}+{\cal O}\left({p^4\over m_h^4}\right)\right]
 \label{a29a}\ee
 \be iG_{{_{SM}},{_{SM}}}(p)=m_{_{SM}}^{2\Delta_{_{SM}}-4}\left[\bar b_0-\bar b_2{p^2\over m_{_{SM}}^2}+{\cal O}\left({p^4\over m_{_{SM}}^4}\right)\right].
\label{a30}\ee
Then,  expanding the propagator as
\be
i\langle \chi \chi\rangle^{-1}(p)\simeq  f_r^2(p^2+m_r^2)+\cdots
\label{a31}\ee
we obtain
\be
f_r^2m_r^2=\bar b_0~m_{_{SM}}^{2(\Delta_{_{SM}}-2)}+{M^8\over \l_0^2 m_h^{2(\Delta_h-2)}\bar a_0}
\sp
f_r^2=-\bar b_2~m_{_{SM}}^{2(\Delta_{_{SM}}-3)}+{M^8\bar a_2\over \l_0^2 m_h^{2(\Delta_1-1)}\bar a_0^2}
\label{a31a}\ee

From now on we will specialize to the case $\Delta_h=\Delta_{_{SM}}=4$ relevant for this paper.
Since we assume that both $m_{h,{_{SM}}}\ll M$ the terms proportional to $M^8$ in (\ref{a31a}) dominate the rest of the terms.
Consequently,
\be
f_r^2={\bar a_0^2\bar a_2\over \l_0^2}{M^8\over  m_h^{6}}-\bar b_2~m_{_{SM}}^{2}\sp m_r^2=\bar a_0 m_h^2\left(1+{\bar b_0\bar a_0\l_0^2\over \bar a_2}{m_h^4m_{_{SM}}^4\over M^8}-{\bar b_1\bar a_0^2\l_0^2\over \bar a_2}{m_h^6 m_{_{SM}}^4\over M^8}+\cdots \right)
\label{a31b}\ee
Therefore,
the renormalization of  $f_a^2$ due to visible theory effects is
\be
{\delta f_a^2\over f_a^2}\sim ~{m_{_{SM}}^2\over m_h^2}\left({m_h\over M}\right)^{8}
\ee
In particular the visible theory contribution to the mass is $~m_{_{SM}}^4$ which should identified with $\Lambda_{QCD}^4$ for the SM. Therefore the second term in the formula for the mass in (\ref{a31b}) is essentially $\Lambda_{QCD}^4/f_a^2$.

However, here the axion has also a mass contribution that originates in the hidden sector and is proportional to the hidden topological susceptibility, $m_h^2$ \cite{Witten:1979vv, Veneziano:1979ec}.
We conclude this section by stressing that the analysis above is valid when $p\ll {\rm min}(m_h,m_{_{SM}})$.
In this regime all higher derivative corrections to the propagating axion are suppressed.
However, the axion mass being mostly $m_h$ in our setup means that phenomenologically, we are interested in the limit $m_h\ll m_{_{SM}}$, and also to see what happens when $m_h\ll p\ll m_{_{SM}}$. We address these issues in the next subsection.

\subsection{Other regimes\label{reg}}

In all of the above we have made no assumption on the ordering of the characteristic dynamical scales $m_h$ and $m_{_{SM}}$ of the hidden theory and the SM respectively. The only assumption was that $p\ll min(m_h,m_{_{SM}})$.
We will now investigate the rest of the space of parameters

$\bullet$ We first assume that $m_h\gg m_{_{SM}}$.
In that case we can investigate one more regime, namely $m_h\gg p\gg  m_{_{SM}}$.
In this regime the $0^{+-}$ glueball of the hidden theory is point-like and featureless while the associated QCD glueball is fat and unstable.
   This is precisely the behavior of a fundamental (non-composite) axion field.
The hidden two-point function has still the form (\ref{a29a}). On the other hand the SM two-point function is instead
\be
iG_{{_{SM}},{_{SM}}}=\langle O_{_{SM}}(p)O_{_{SM}}(-p)\rangle=p^4\log{p^2\over m_{_{SM}}^2}\left[\hat b_0+\hat b_2{m_{_{SM}}^2\over p^2}+{\cal O}\left({m_{_{SM}}^4\over p^4}\right)\right]\sp p\gg  m_{_{SM}}
\label{a29b}\ee
 In this case, $\l^2 G_{{_{SM}},{_{SM}}}(p)\ll {1\over G_{h,h}(p)}$ and the SM corrections to the axion kinetic data are tiny.
Therefore, in this regime the axion data are given by (\ref{a28}) and are determined by the scales of the hidden theory as well as the messenger scale.

$\bullet$ We now investigate the opposite limit, $m_h\ll m_{_{SM}}$.
In that case we can investigate one more regime, namely $m_{_{SM}}\gg p\gg  m_h$.
 Here the $0^{+-}$ glueball of the hidden theory is fat  while  the associated QCD glueball is point-like (and unstable).
   This is clearly distinct from the behavior of a fundamental (non-composite) axion field.
 The SM two-point function is given again by (\ref{a30}) but now the hidden two-point function becomes
\be
G_{h,h}(p)=p^4\log{p^2\over m_h^2}\left[-\hat a_0+\hat a_2{m_h^2\over p^2}+{\cal O}\left({m_h^4\over p^4}\right)\right]\sp p\gg  m_h
\label{a29c}\ee
with $\hat a_0,\hat a_2$ dimensionless numbers of order ${\cal O}(1)$.

In this regime the  $\chi$ propagator is
\be
\langle \chi\chi\rangle^{-1}\simeq {M^8\over \l_0^2 p^4\log {p^2\over m_h^2}\left[\hat a_0+{\cal O}\left({m_h^2\over p^2}\right)\right]}+m_{_{SM}}^4\left[\bar b_0+{\cal O}\left({p^2\over m_{_{SM}}^2}\right)\right]
\label{a29d}\ee

To ascertain which term dominates we will define the scale
\be
\mu\equiv {M^2\over m_{_{SM}}}
\label{b42}\ee

We always have $\mu\gg m_{_{SM}}$. In this case, for all $m_h\ll p\ll m_{_{SM}}$,  the $G^{-1}_{h,h}$ part in (\ref{a29d}) dominates and
\be
   \langle \chi\chi\rangle^{-1}\sim {M^8\over  p^4\log {p^2\over m_h^2}}
\label{b45}   \ee
over the whole energy regime $m_h\ll p\ll m_{_{SM}}$.

This is a non-standard non-local axion kinetic term  that in configuration space behaves as
\be
  \langle \chi\chi\rangle^{-1}\sim {M^8}\log|x|
\label{b46}  \ee
  and in configuration space the quadratic term for the axion is (up to factors of order ${\cal O}(1)$
  \be
S_{eff}\simeq{M^8\over 2}\int d^4x_1d^4x_2~\chi(x_1)\log{|x_1-x_2|\over m_h}\chi(x_2)+\int d^4x~\chi(x) O_{_{SM}}(x)
\label{b47}\ee
It should be stressed that such an unusual (highly non-local) quadratic term is valid for distances $m_{_{SM}}^{-1}< \ell<m_h^{-1}$.

Despite this unusual feature the interaction induced by such a term is weak.
As the kinetic term is non-standard, we will characterize the strength of the interaction by the values of $\langle \chi\chi\rangle$ that controls the interaction between sources coupled to axions
\be
{1\over M^4}\left({m_h\over M}\right)^4 \precsim \langle \chi\chi\rangle  \precsim {1\over M^4}\left({m_{_{SM}}\over M}\right)^4
\label{b47a}\ee
Note that the analogous value for the $ \langle \chi\chi\rangle$ in the low energy regime is $m_h^4/M^8$.
The momentum dependence though is non-standard and maybe measurable. As $m_{_{SM}}\sim \Lambda_{QCD}$ and $m_h$ can be made much lighter, such a case must be analyzed from first principles in order to decide the experimental constraints on $m_h$.

In the case where the hidden theory is a CFT, $m_h=0$.
The physics in this case can be established by taking the $m_h\to 0$ limit in the case discussed above.
The scale $\mu$ can again be defined as in (\ref{b42}).

$\bullet$ Finally the only remaining regime is $M\gg p\gg max(m_h,m_{_{SM}})$. In this regime both glueballs are fat and the couplings are expected to be non-local.
Indeed, both $G_{h,h}$ and $G_{{_{SM}},{_{SM}}}$ are given by (\ref{a29c}) and (\ref{a29b}), from (\ref{a29}) we obtain
\be
\langle\chi\chi\rangle^{-1}={M^8\over \l_0^2 G_{h,h}}-G_{{_{SM}},{_{SM}}}=-{M^8\over \l_0^2 p^4\log {p^2\over m_h^2}\left[\hat a_0+\cdots\right]}-p^4\log{p^2\over m_{_{SM}}^2}\left[\hat b_0+\cdots\right]
\label{b49}\ee
 In this case the first term in (\ref{b49}) dominates and we obtain
\be
\langle\chi\chi\rangle^{-1}\sim {M^8\over N^2\l_0^2 p^4\log {p^2\over m_h^2}}
\label{b50}\ee
In this regime, the axion kinetic term is similar to the one in (\ref{b47}).

As expected, the interaction is non-local as in this regime the relevant glueballs are of finite size.
The effective strength of the interaction, induced on sources coupled linearly to the axion with strength one is proportional to $\langle \chi\chi\rangle$ and (neglecting logs) varies between
\be
{{\rm max}(m_h,m_{_{SM}})^4\over M^8}\leq \langle \chi\chi\rangle \leq {1\over M^4}
\label{b52}\ee

We note that from (\ref{b47a}) and (\ref{b52}) the effective strength of the interaction is increasing as we move towards the UV.

 We conclude our discussion as follows: in the general case where we have multi-scalar couplings between various scalar operators of the two theories, generically these lead to emergent interactions via scalars that are very heavy (and therefore not very relevant for low energy physics) as their masses are at the  messenger scale $M$.
However,  the instanton density\footnote{There may be several instanton densities in the hidden theory, as there are in the SM model.} of the hidden theory gives rise to an emergent axion-like field that couples (weakly in most cases) to the SM model instanton densities\footnote{And may also couple to other CP-odd gauge invariant operators of the SM model.}.

\subsection{Other relevant axion-generating operators\label{eta}}

Although our discussion so far has involved scalar operators with special properties, we have focused on the case of the instanton densities.
In the relevant literature that includes  composite axions, \cite{Kim2}, the canonical axions are related to (light) meson fields and the associated (chiral) symmetries acting on fermions.

In gauge theories with fundamentals, global symmetries can be carried by both bosons and fermions. It is however some of the symmetries of the fermions that have ($U(1)_A$) anomalies and therefore provide appropriate couplings to the QCD Instanton density.
In most of the literature, one postulates a hidden gauge theory (as we do here) with a $\Lambda$ scale that is $\gg \Lambda_{QCD}$ and at least two fermionic U(1)'symmetries (ie. two sets of fermions) are needed to generate an axion. The reason is that with only one, the only meson is the $\eta'$ and this is heavy because  the $U(1)_A$ gives it a mass of order\footnote{There is a loophole in this argument: at large enough $N_c$ but finite number of flavors, the $\eta'$ mass is suppressed as ${N_c^{-1}}$.} $\Lambda$.
With two symmetries, the spectrum can be adjusted so that one of the mesons gets a mass only from $U(1)_{A}$ of QCD and therefore be relatively light, \cite{Kim2}.

Here we entertained a different possibility: that the hidden gauge theory is connected to the SM at some very large scale via irrelevant couplings and therefore we can allow its characteristic scale to be very low, and in particular much lower that $\Lambda_{QCD}$. This we applied  to the hidden instanton density (that becomes the axion) but it can also be applied
 to more complicated pseudo-scalar operators (generalized pions and $\eta$'s)
 that allow many more possibilities in terms of couplings and masses.

The analysis however can be extended by adding the possibility of $\eta'$ (scalar) operators in the presence of fermionic fundamental or other representations.
In a large-N$_c$ ``regular" theory,  color representations can have at most $N_c^2$ degrees of freedom. Therefore, the fermion representations allowed are
$\Yboxdim8pt\yng(1)$, ${\Yboxdim8pt\yng(2)}$, ${\Yboxdim8pt\yng(1,1)}$ and their conjugates as well as the adjoint.
All of them carry fermionic symmetries, out of which the $U(1)_A$ ones are anomalous. Because of this the associated $\eta'$ particles (one for each color symmetry)
 are massive and their masses do not satisfy the GOR relations, but they scale as $m_{\eta'}\sim N_f{\Lambda \over N_c}$ for the fundamentals and $m_{\eta'}\sim N_f{\Lambda}$ for all other representations.
In the presence of more than one sets of fermions, linear combinations can be made free of the $U(1)_A$ symmetry of the strongest of the gauge groups and therefore arrange so that the relevant $\eta'$ to be lighter than the heavier of the gauge theory scales as in \cite{Kim2}.
The proper discussion of such cases in this framework involves more than one pseudo scalar operators and scalings that are subtler than the generic ones presented in section \ref{IR}.

In particular, in the case of light quarks with mass $m_q$ in the SM,  the scales $b_0$ and $b_2$ become
\be
b_0=\Lambda_{_{SM}}^3m_q\sp  b_2\sim \Lambda^2_{_{SM}}
\ee
where now the operator is the standard $\eta'$ of the SM.

For the case where the bifundamental messengers have light masses, $M\ll \Lambda_h,\Lambda_{_{SM}}$, as discussed in section \ref{strong},  the $\eta'$ associated with the messengers has a mass $\sim {max(\Lambda_h,\Lambda_{_{SM}})\over N_c}$ where $N_c$ is the number of colors of the hidden gauge group. In the presence of extra quarks charged under the hidden gauge group there is a linear combination of $\eta'$ that has a mass $\sim {min(\Lambda_h,\Lambda_{_{SM}})\over N_c}$.

  We will not pursue further these possibilities. Some  have already been discussed in the literature, \cite{Kim2}.

\section{Higher Interactions}\label{Interactions}

As we discussed in section~\ref{CoupledSectors}, we assumed a perturbative structure for the two theories $T_{1,2}$ and treated both of them in the quadratic approximation. This was implemented in~(\ref{b2})-(\ref{a10}). The only interaction present was the $O_1 O_2$ deformation coupling the two sectors.

One can further extend the analysis to the non-linear regime by considering interactions both inherently present in each of the uncoupled sectors as well as further cross-interactions between them. The functional
(\ref{b1}) contains in principle all the possible self-interactions of each sector so in that sense it is exact. To accommodate higher point cross-interactions one can deform it with terms such as $\lambda_{1122} \int d^4 x \, O_1 O_1 O_2 O_2$ etc. In the rest of this section we will reserve the capital Latin indices $I, J, K , L$ as indicators of the two theories, hence they can take the two values $1,2$. As an example a generic four-operator interaction will take the form
\be
\lambda_{IJKL} \int d^4 x \, O_I O_J O_K O_L\;.
 \ee
 One would then have to expand the total functional in powers of the external sources to obtain the appropriate correlator. This can be performed in a perturbative expansion in powers of the various couplings. The expansion is around the Gaussian-approximation resulting in (\ref{b3}). Instead of listing all the possible diagrams, it is convenient to rewrite the functional (\ref{b3}) as
\bea
e^{i W(J_1, J_2)} &=& N^{-1} \int \mathcal{D} \phi_1 \mathcal{D} \phi_2 e^{i S_G(\phi_I)}\, , \nn \\
S_G(\phi_I) &=& \int {d^4 p\over (2\pi)^4}\left[{1\over 2}\left(\begin{matrix}\phi_1, & \phi_2\end{matrix}\right)\left(\begin{matrix}{1\over G_{11}(p)}& -{\l}\\
-{\l}& {1\over G_{22}(p)}\end{matrix}\right)\left(\begin{matrix}\phi_1\\ \phi_2 \end{matrix}\right)+J_1 \phi_1 +J_2 \phi_2 \right] \nn \\
\label{d1}
\eea
and add the interactions by deforming the functional $S_q(\phi_I)$ with the following terms
\bea\label{d2}
S_{int}[\phi_I] = \int \prod_{i=1}^3 d^4 p_i ~\delta \left(\sum_{j=1}^3 p_j \right)\left[ \frac{V_{I J K}^{(3)}(p)}{3!} \phi_I(p_1)\phi_J(p_2)\phi_K(p_3)
\right] \nn \\
  + \int \prod_{j=1}^4 d^4p_j ~\delta \left(\sum_{i=1}^4 p_i \right)\left[ \frac{V^{(4)}_{I J K L}(p)}{4!} \phi_I(p_1)\phi_J(p_2)\phi_K(p_3)\phi_L(p_4) \right]+\cdots
\eea
This ansatz encapsulates both the interactions present in each uncoupled theory itself as well as cross interactions.
The advantage of this rewriting is that one can easily take into account quantum corrections at any loop order, the couplings $V^{(3)}_{I J K}$ at tree level correspond to the bare couplings $\lambda_{I J K}$, while quantum corrections renormalise them\footnote{We also note that $\Gamma[\phi_I] = S_G[\phi_I] + S_{int}[\phi_I]$ can be thought of as the effective action for the coupled theories computed at a given loop order, for more details see~\cite{Cornwall:1974vz}.}.
One such effect of interactions is to renormalise the two-point function via loop corrections. This is the main effect we will be interested in, since it affects the propagator of the emergent (pseudo) scalar (\ref{a16}) as shown in (\ref{d5}) below. In particular loop corrections cause a shift (or renormalization) in the quadratic matrix part of $S_G$ of the form
\be\label{d3}
\left[G^{-1}\right]^{ren}_{I J} = \left(\begin{matrix}{1\over G_{11}(p)} + \Sigma_{11}(\Lambda, p) & -{\l} + \Sigma_{12}(\Lambda, p) \\
-{\l} + \Sigma_{21}(\Lambda, p)& {1\over G_{22}(p)} + \Sigma_{22}(\Lambda, p)\end{matrix}\right) \, ,
\ee
where with $\Sigma_{I J}(\Lambda, p)$ we denote the matrix elements of the loop corrections, that generically depend on the momentum $p$, but also on any cutoff scale which we denote by $\Lambda$. In particular for us such a cutoff will be set by the messenger scale and therefore $\Lambda = M$. In Appendix~\ref{Intloops} we describe this computation and provide explicit expressions for the matrix elements $\Sigma_{I J}(p, \Lambda)$ in terms of integrals, considering quartic and cubic interactions. Their physical properties will be analysed below in different regimes.

To obtain the dressed correlators, one can invert the renormalised matrix (we assume that $\Sigma_{I J}$ is symmetric) to obtain
\be\label{d4}
G^{ren}_{I J} = \left(
\begin{array}{cc}
 \frac{\Sigma_{22}+\frac{1}{G_{22}}}{\left(\Sigma_{11}+\frac{1}{G_{11}}\right)
   \left(\Sigma_{22}+\frac{1}{G_{22}}\right)-\left(\Sigma_{12}- \lambda \right)^2} &
   \frac{\lambda-\Sigma_{12}}{\left(\Sigma_{11}+\frac{1}{G_{11}}\right)
   \left(\Sigma_{22}+\frac{1}{G_{22}}\right)-\left(\Sigma_{12}-\lambda\right)^2} \\
 \frac{\lambda-\Sigma_{12}}{\left(\Sigma_{11}+\frac{1}{G_{11}}\right)
   \left(\Sigma_{22}+\frac{1}{G_{22}}\right)-\left(\Sigma_{12}-\lambda\right)^2} &
   \frac{\Sigma_{11}+\frac{1}{G_{11}}}{\left(\Sigma_{11}+\frac{1}{G_{11}}\right)
   \left(\Sigma_{22}+\frac{1}{G_{22}}\right)-\left(\Sigma_{12}-\lambda \right)^2} \\
\end{array}
\right)
\ee
One can now match the $G^{ren}_{22}$ element of the correlator that corresponds to $\langle O_2 O_2 \rangle^{ren}$ with the expression for the corrected correlator (\ref{a17}) that arises from the presence of a dynamical ``emergent" field with kinetic operator $K(p)$, see (\ref{a16}). This identifies the latter as follows (up to the messenger cutoff scale $M$)
\bea\label{d5}
K(M, p) &=& g^2 \frac{\Sigma_{11}(M, p) + \frac{1}{ G_{11}(p)}}{(\Sigma_{12}(M, p) - \lambda )^2 - \Sigma_{22}(M, p)(\Sigma_{11}(M, p) + \frac{1}{G_{11}(p)})} \, , \nn \\
g^2 \langle \chi(p) \chi(-p) \rangle &=& g^2 K^{-1}(M,p) = - \Sigma_{22}(M, p) + \frac{(\Sigma_{12}(M, p) - \lambda)^2}{\Sigma_{11}(M, p) + \frac{1}{G_{11}(p)}} \, , \nn \\
&\approx & - \Sigma_{22}(M, p) + \lambda^2 G_{11}(p)\left(1 - 2 \frac{\Sigma_{12}(M,p)}{\lambda} - G_{11}(p) \Sigma_{11}(M,p)  \right)\, , \nn \\
\eea
where in the last line we assumed that the renormalization effects are perturbative in nature and therefore the matrix elements $\Sigma_{I J}$ are parametrically small and we can keep the leading correction.
With this identification we can now study the effects that interactions have in interpreting the effects of the ``hidden" theory $T_1$ as coming from some dynamical ``emergent" field coupled linearly to the operator $O_2$ of $T_2$ beyond the Gaussian regime of the previous section. As an example, we will now discuss in detail the physical implications of cubic and quartic corrections in several regimes of interest.

\subsection{One-loop correction to the propagator due to quartic interactions}\label{Quartic}

We now consider quartic interactions and find their correction to the matrix correlator. The procedure involving the computation of one loop diagrams is discussed in Appendix~\ref{Quarticloop}. We use the two-point function (\ref{a10}) to compute the one loop diagrams (\ref{1loop4v})with a cutoff method. We will use Latin indices $I,J,K,L = 1,2$ to label the fields $\phi_1,  \phi_2$. The simplest case for the four-point vertex is the isotropic one for which $V^{(4)}_{I K L J} = V_4 \left( \delta_{I K} \delta_{L J} + \delta_{I L} \delta_{K J}\right)$. From the relevant Feynman graphs, thanks to momentum conservation at the four-vertex, it is easy to see that there cannot be any external momentum dependence and thus $\Sigma^{(4)}_{I J}(M)$ can depend only on the cutoff $M$.

To perform the computation one has to use the matrix propagator (\ref{a9}) with
\be
G_{11}(p) \sim p^{2 \Delta_1 - 4}\sp G_{22}(p) \sim p^{2 {\Delta_2} - 4}\;.
 \ee
 As described in the Appendix~\ref{Quarticloop}, all the matrix elements of $\Sigma^{(4)}_{I J}$ can be computed in terms of the function
$I(a,b,A,B)$ (\ref{mainint2}). The result is given in (\ref{B2}) and admits the following expansion in powers of the messenger scale cutoff $M$
\be
\Sigma^{(4)} = V_4 \begin{pmatrix}
A_{11} \left(\frac{ M^{8-2 \Delta_1 } }{\lambda^2(8-2 \Delta_1)} + \frac{4 M^{16- 2 \Delta_2 - 4 \Delta_1}   }{\lambda^4(16- 2 \Delta_1 - 4 \Delta_2)}  + ... \right)  & A_{12} \left(\frac{ M^{4}}{4 \lambda}   + \frac{16  M^{12 - 2( \Delta_1+ \Delta_2)} }{\lambda^3(12 - 2( \Delta_1+ \Delta_2))}  + ... \right)  \\
A_{12} \left(\frac{ M^{4}}{4 \lambda}   + \frac{16  M^{12 - 2( \Delta_1+ \Delta_2)} }{\lambda^3(12 - 2( \Delta_1+ \Delta_2))}  + ... \right) & A_{22}  \left(\frac{ M^{8-2 \Delta_2 } }{\lambda^2(8-2 \Delta_2)} + \frac{4  M^{16- 2 \Delta_1 - 4 \Delta_2}    }{\lambda^4(16- 2 \Delta_2 - 4 \Delta_1)}  + ...  \right)
\end{pmatrix}
\label{CFT-CFT}
\ee
where the precise constant numerical coefficients $A_{I J}$ can be found using (\ref{mainint2}) together with
(\ref{B2}). One notices that upon expressing the couplings in terms of dimensionless parameters
\be
\lambda = \lambda_0 M^{4- \Delta_1 -  \Delta_2}\sp V_4 =V_4^0 M^{4- 2 \Delta_1 - 2  \Delta_2}\;,\ee
the matrix elements scale as expected from dimensional analysis in (\ref{d3}). In particular the leading correction is
\be
\Sigma^{(4)} = V_4^0 \begin{pmatrix}
A_{11} \frac{ M^{4-2 \Delta_1 } }{\lambda_0^2(8-2 \Delta_1)} + ... & A_{12} \frac{  M^{4-\Delta_1 -  \Delta_2}}{4 \lambda_0} + ...  \\
A_{12} \frac{  M^{4-\Delta_1 -  \Delta_2}}{4 \lambda_0} + ...    & A_{22} \frac{ M^{4-2 \Delta_2 } }{\lambda_0^2(8-2 \Delta_2)} + ...
\label{CFT-CFT2}
\end{pmatrix}
\ee
In the case that one or the two theories develops a mass gap $m$ in the IR, the leading cutoff dependence still keeps the same form, up to corrections in the form of powers of the dimensionless ratio $m / M$.
\\
\\
We note that even though the correction matrix $\Sigma_{I J}^{(4)}$ is momentum independent, the momentum dependence will come from inverting the matrix in order to compute the propagator as shown in passing from (\ref{d3}) to (\ref{d4}). In addition it is easy to see that for integer $\Delta_1, \Delta_2$ one can get logarithmic divergences. The structure of the integrals then is that given in (\ref{quarticlogarithmic}). We will describe such cases with more detail in section~\ref{Quarticregimes}.

\subsection{Implications for the emergent axion propagator}\label{Quarticregimes}

We now work out the implications of quartic interactions to the emergent axion propagator.
The matrix $K(p)$ given by (\ref{d5}) is going to be identified as the inverse propagator of an emergent field that we integrate in, see (\ref{a16}). This emergent field captures the dominant physical effects of the theory $T_1$ which we refer to as the hidden sector. Since we work in perturbation theory we can only trust small corrections around the non-perturbative quadratic result given in (\ref{a22}).  We will again re-express the coupling constants in terms of dimensionless parameters as $m= m_0 M, \, \lambda = \lambda_0 M^{4- \Delta_1 -  \Delta_2}$ and $V_4 =V_4^0 M^{4- 2 \Delta_1 - 2  \Delta_2}$. The coupling $g$ for the emergent field $\chi$ in case that this field is identified with an emergent axion can be conveniently expressed as $g= N M^{4 - \Delta_2}$ to make the axions dimensionless. All the constants in this section will depend on these dimensionless parameters.  We also notice that $\Sigma_{12}\sim M^{4-\Delta_1 -  \Delta_2}\, , \Sigma_{11}\sim M^{4- 2 \Delta_1} \, , \Sigma_{22}\sim M^{4 - 2  \Delta_2}$ which are the expected canonical dimensions of these elements that shift the original inverse propagator elements appropriately. We also recall that $\Delta_1$ is the dimension of the hidden sector operator  while $\Delta_2$ refers to the dimension of the SM operator.
\\
\\
Using (\ref{d5}) together with (\ref{CFT-CFT2}) we find
\be
\langle \chi(p) \chi(-p) \rangle_{1-loop} \approx  - \frac{a_0 V_4^0}{g^2  M^{ 2 \Delta_2 - 4}} + \frac{\lambda_0^2 p^{2 \Delta_1 - 4}}{g^2  M^{ 2\Delta_1 + 2 \Delta_2  - 8}} \left(1 -  V_4^0 \left[ \frac{b_0}{\lambda_0} + c_0 \frac{p^{2 \Delta_1 - 4}}{M^{2 \Delta_1 - 4}}  \right] \right)
\label{cutoffcorrelator}
\ee
where $a_0, b_0, c_0$ are numerical constants and $V_4^0 \ll 1$ so that the perturbative approximation is trustworthy. The first term depends only on the cutoff and is either to be subtracted or vanishes in the limit of large cutoff $M$ (i.e. with $g = N M^{4 - \Delta_2}$  it is found to vanish as $M^{-4}$). The $b_0$ term results in a constant wavefunction renormalization. The momentum dependent term shifts the quadratic solution (\ref{a22}) by a small amount and therefore provides the most interesting effect. Let us now list the following cases depending on the conformal dimensions of the operators:
\begin{itemize}
\item  The relevant operator case $ \Delta_1 < 2$.

In this case the $c_0$ term deformation is relevant and shifts perturbatively the pole of the propagator from $p = 0$.

\item At $\Delta_1 = 2$ one finds a logarithmic scaling for the correction in terms of the cutoff $M$
\be
\langle \chi(p) \chi(-p) \rangle_{1-loop} \approx  - \frac{a_0 V_4^0}{g^2  M^{ 2 \Delta_2 - 4}} + \frac{\lambda_0^2 \log(p/M)}{g^2  M^{ 2 \Delta_2  - 4}} \left(1 -  V_4^0 \left[ \frac{b_0}{\lambda_0} + c_0 \log(p/M) \right] \right)
\ee

\item Beyond that ($\Delta_1 > 2$), one finds that the deformation is irrelevant and does not affect the non-perturbative quadratic solution in the IR. Therefore operators with $\Delta_1 = 4$ do lead to perturbative corrections to the quadratic result (\ref{a22}) except from an overall wave-function renormalization.

\item A special case is given by a standard model operator of dimension $ \Delta_2 = 4$. In such a case it is easy to see from (\ref{CFT-CFT2}) that the matrix elements $\Sigma_{22}, \, \Sigma_{12} \rightarrow 0$ which leads to $a_0 = b_0 = 0$ i.e. no wavefunction renormalization. If furthermore $\Delta_1 > 2$ (as in our case for a ``hidden'' instanton density $\Delta_1 = 4$), the effects of such a combination of operators are exactly captured by the quadratic result.

\end{itemize}

We therefore conclude that operators with $\Delta = 4$ are special in that they are protected by UV effects to one loop order (assuming quartic interactions) that could effect both an overall wavefunction renormalization and/or a shift in the poles of the propagator of the non-perturbative Gaussian treatment. We will now proceed to the study of cubic interactions.

\subsection{One loop corrections due to a cubic vertex}\label{Cubic}

In this section we will repeat the analysis performed for the quartic case~\ref{Quartic}, now for the cubic vertex.
The relevant computation is presented with more detail in appendix~\ref{Cubicloop}. One notices two important differences with the quartic case. The one loop correction scales as $(V_3)^2$ since one needs two cubic vertices to form a one loop graph, and the matrix $\Sigma_{I J}^{(\text{1-loop) (3)}}(p, M)$ now depends both on the messenger scale cutoff $M$ and the momentum $p$.
In addition for dimensional reasons it is found than in a $p/M$ expansion, the leading term is momentum independent and scales accordingly to the dimension of the related matrix element of the inverse correlator~(\ref{a9}) similarly to what happens in the quartic case (\ref{CFT-CFT}). The corrections can then be organised in an expansion $\sum_{n} a_{n} (\frac{p}{M})^{2n}$ where the momentum can appear only in even powers due to rotational invariance of the integrals. We can directly proceed analysing the physical regimes and properties of the cubic corrections.

\subsection{Implications for the emergent axion propagator}\label{Qubicregimes}

We will now discuss in more detail the physical regimes of the cubic corrections in a similar spirit to~\ref{Quarticregimes}
We will again use the coupling constants scaling in terms of the bare constants as $m= m_0 M, \, \lambda = \lambda_0 M^{4- \Delta_1 -  \Delta_2}$ and $V_3 =V_3^0 M^{4- 2 \Delta_1 - 2  \Delta_2}$. All the arbitrary constants in this section will depend on these dimensionless parameters.

Using the results of the appendix~\ref{Cubicloop}, we find that the matrix elements $\Sigma_{I J}^{(3)}$  scale as
\be
\Sigma_{12}^{(3)} =  \left(V_3^0 \right)^2 \begin{pmatrix}
B_{11} M^{4-2 \Delta_1 } \left(1 + C_{11} \frac{p^2}{M^2} + ... \right) & B_{12}  M^{4-\Delta_1 -  \Delta_2} \left(1 +  C_{12} \frac{p^2}{M^2} + ... \right) \\
B_{12} M^{4-\Delta_1 -  \Delta_2} \left(1 + C_{12} \frac{p^2}{M^2}   + ... \right)    & B_{22}  M^{4-2 \Delta_2 }  \left(1 +  C_{22} \frac{p^2}{M^2}  + ... \right)
\label{cubicsigma}
\end{pmatrix}
\ee
with $B_{I J}, C_{I J}$ numerical coefficients. These matrix elements take the expected scaling and are presented in a perturbative fashion for small momenta. Notice again that $\Delta_1$ is the hidden operator dimension while $ \Delta_2$ refers to the SM operator.
\\
\\
From these elements, we obtain the cubic corrected propagator of the emergent field $\chi$ up to first order in $p/M$
\bea
\langle \chi(p) \chi(-p) \rangle  &\sim &  - \frac{a_0 \left(V_3^0\right)^2}{g^2  M^{ 2 \Delta_2 - 4}}\left(1 + \frac{ a_1 p^2}{M^2} \right) \nn \\
&+& \frac{\lambda_0^2 p^{2 \Delta_1 - 4}}{g^2  M^{ 2\Delta_1 + 2 \Delta_2  - 8}} \left(1 -  \left(V_3^0\right)^2 \left[ \frac{b_0}{\lambda_0} + \frac{b_1 p^2}{\lambda_0 M^2} + \frac{ c_0 p^{2 \Delta_1 - 4}}{M^{2 \Delta_1 - 4}} + \frac{  c_1 p^{2 \Delta_1 - 2}}{M^{2 \Delta_1 - 2}} \right] \right) \nn \\
\label{cutoffcorrelatorcubic}
\eea
with $a,b,c$'s numerical constants.
The results of the previous section~\ref{Quarticregimes} apply with the following extra modifications:
\begin{itemize}
\item There are now further higher derivative terms compared to the quartic case parametrised by $a_1, b_1, c_1$. All these terms are irrelevant in the IR compared to the leading terms and suppressed by powers of the cutoff.

\item There is the possibility of new poles (an infinite number of them) arising from all these higher derivative terms. Since to fully clarify such a possibility would require computations up to the cutoff our perturbative approach does not provide a systematic and consistent method to analyse such physical effects. A better approach to study such an infinite number of resonances based on holography is described in section~\ref{Hologaxion}.

\end{itemize}

By studying the leading terms, the conclusions of the previous section remain unaltered. In particular
for $ \Delta_2 = 4$ we get that the $a$ and $b$-terms vanish completely. The $c$-terms that are left are found to be irrelevant for $\Delta_1 > 2$  or in other words the effects of ``hidden'' fields vanish for such operators. These conclusions are in line with the previous sections. Operators of conformal dimension $\Delta = 4$ do not get any corrections with cutoff dependence even after including perturbative interactions.

\section{The holographic axion}\label{Hologaxion}

We  now investigate the special case where the hidden theory $T_1$ is a large $N$ holographic theory.

The general action can be written as
\be
S=S_{1}+S_{12}+S_2
\ee
where the interaction term has been defined in (\ref{a1}), $S_1$ is the action of the holographic theory, and $S_2$ the action of the SM.
Applying the holographic correspondence, we can write\footnote{For a scalar operator $O_\Delta(x)$ of dimension $\Delta$ dual to a field $\Phi_M(x,z)$ of mass $(M\ell)^2= \Delta(\Delta-4)$ the asymptotic behaviour would be
$\Phi_M(x,z)\approx z^{\Delta-4}O_\Delta(x)$.}
\be
\langle e^{iS_{12}}\rangle=\int_{\lim_{z\to 0}a(x,z)= O_2(x)} {\cal D}a~e^{iS_{\rm bulk}[a]}
\label{e1}\ee
where on the left, the expectation value is taken in the holographic theory $T_1$.  $S_{\rm bulk}[a]$ is the bulk gravity action, $z$ is the holographic coordinate, $a$ is the bulk field dual to the operator $O_1$ of dimension $\Delta=4$ and the gravitational path integral has boundary conditions for $a$ to asymptote to the operator $O_2$ near the AdS boundary. We have also neglected the other bulk fields.

By inserting a functional $\delta$-function  we may rewrite (\ref{e1}) as
\be
\langle e^{iS_{12}}\rangle=\int_{\lim_{z\to 0}a(x,z)= \phi(x)} {\cal D}a(x,z){\cal D}\phi(x){\cal D}k(x)~e^{iS_{\rm bulk}[a]+i\int k(x)(\phi(x)-O_2(x))}
\label{e2}\ee
If we now integrate $\phi(x)$ first in the path integral transform, we obtain the Legendre transform of the Schwinger functional of the bulk axion which becomes the bulk effective action. This corresponds in holography to switching boundary conditions at the AdS boundary from Dirichlet to Neumann, and where $k(x)$ is the expectation value of the operator $O_1$.
We finally obtain
\be
\langle e^{iS_{12}}\rangle=\int_{\lim_{z\to 0}\partial_za(x,z)=z^3 k(x)} {\cal D}a(x,z){\cal D}k(x)~e^{iS_N[a]-i\int k(x)O_2(x)}
\label{e22}\ee
We may imagine the SM action as coupled at the radial scale $z_0\sim 1/M$ to the bulk action.
Following holographic renormalization \cite{Bianchi:2001kw, Bianchi:2001de}, we may then rewrite the full bulk+brane action of the emergent axion as
 \be
S_{total}=S_{bulk}+S_{brane}
\label{e3}\ee
\be
S_{bulk}=M_P^3\int d^5x\sqrt{g}\left[Z(\pa a)^2+{\cal O}((\pa a)^4)\right]
\label{e4}\ee
\be
S_{brane}=\delta(z-z_0)\int d^4x\sqrt{\gamma}\left[ \lambda \hat a(x)O_2(x)+M^2(\pa \hat a)^2-\Lambda^4 \hat a^2+\cdots\right]
\label{b5}\ee
where $\hat a(x)\equiv a(z_0,x)$ is the induced axion on the brane.
As we will be interested at energies $E\ll M$ we can ignore higher axion terms like $a\square ^2 a$ on the brane.

 In the boundary action (\ref{b5}) $\gamma$ is the induced four-dimensional metric. The first term in (\ref{b5}) is the coupling of the axion to the SM Instanton densities descending from (\ref{a1}). The kinetic and mass terms of the axion in the brane action come from the quantum effects of the SM fields, as explained at the end  of section \ref{axion}.
The ellipsis stands for the rest of the SM action as well as higher derivative corrections to the brane axion field action.

In the bulk action, (\ref{e3})  we have neglected the graviton and other scalar fields dual to other scalar operators of the ``hidden" holographic theory $T_1$. The factor $Z$ in (\ref{e3}) in general depends on the various other scalars fields. For the case of holographic YM this action has been studied in detail in \cite{ihqcd,cs}. The graviton also couples to the SM action and provides emergent gravity, \cite{bbkn}. Importantly, there is a bulk potential for the
axion but it is due to instantons and therefore it is exponentially suppressed at large N. We have therefore neglected it. To all orders in 1/N,  the bulk axion
has only derivative interactions.
Finally the boundary conditions for the bulk  action are Neumann.
It should be noted that what we have here is a close analogue of the DGP mechanism, \cite{DGP}, with two differences: here we have an axion and also the bulk data are non-trivial.

The main difference in the physics of an emergent axion originating in a holographic theory is that due to the strong coupling effects there is an infinity of axion-like resonances coupled to the SM instanton densities. They correspond to the poles of the two-point function of the instanton density of the ``hidden" holographic theory.
If the holographic theory is gapless, then there is a continuum of modes and as mentioned earlier in such a case the induced axionic interaction is non-local.
If the theory has a gap as large-N YM then there is a tower of nearly stable states at large N that are essentially the 0$^{+-}$ glueball trajectory and act as the KK modes of the bulk axion that couple with variable strengths to the SM instanton densities.

To investigate these interactions we  analyze the propagator of the axion on the SM brane.
To do this we introduce a $\delta$-function source for the axion on the brane and we  solve the bulk+brane equations in the linearized approximation, assuming a trivial profile for the bulk axion\footnote{ This will be the case is the $\theta$-angle of the hidden QFT vanishes. In the jargon of holographic renormalization this is called an `inert' scalar
\cite{Bianchi:2001kw, Bianchi:2001de, Bianchi:2000sm, Bianchi:2003bd, Bianchi:2003ug}.}  while the metric and other scalars have the holographic RG flow profile of a Lorentz-invariant QFT, namely
 \be
 ds^2=dz^2+e^{2A(z)}dx_{\m}dx^{\m}\sp Z(\Phi_i(z))
 \label{b6}\ee
The calculation follows similar ones in \cite{Bianchi:2000sm}-\cite{Bianchi:2003ug} which we reproduce here,
\be
M_P^3\left[\pa_z^2+\left({Z'\over Z}+4A'\right)\pa_z -e^{-2A}\square_4\right]G(x,z)+
\label{b7}\ee
$$
+\delta(z-z_0)(M^2\square_4 -\Lambda^4)G(x,z)=\delta(z-z_0)\delta^{(4)}(x)
$$
where we work in Euclidean 4d space and primes stand for derivatives with respect to $z$. We Fourier transform along the four space-time dimensions to obtain
\be
M_P^3\left[\pa_z^2+\left({Z'\over Z}+4A'\right)\pa_z -e^{-2A}p^2\right]G(p,z)-\delta(z-z_0)(M^2p^2 +\Lambda^4)G(p,z)=\delta(z-z_0)
\label{b8}\ee
where $p^2=p^ip^i$ is the (Euclidean) momentum squared. Later on we will also use $p=\sqrt{p^2}$.

 To solve this, we must first solve this equation for $z>z_0$ and for $z<z_0$
 obtaining two branches of the bulk propagator, $G_{IR}(p,z)$ and $G_{UV}(p,z)$ respectively. The IR part, $G_{IR}(p,z)$ depends on a single multiplicative integration constant as the regularity constraints in the interior of the bulk holographic geometry fix the extra integration constant.
$G_{UV}(p,z)$ is defined with Neumann boundary conditions at the AdS boundary and depends on two integration constants. In the absence of sources and fluctuations on the SM brane, the propagator is continuous with a discontinuous $z$-derivative at the SM brane\footnote{For Randall-Sundrum branes this condition is replaced by $G_{UV}(p,z-z_0)=G_{IR}(p,z_0-z)$, which identifies the UV side with the IR side. This corresponds to a cutoff holographic QFT in the bulk.}
\be
G_{UV}(p,z_0;z_0)=G_{IR}(p,z_0;z_0)\sp \partial_zG_{IR}(p,z_0;z_0)-\partial_zG_{UV}(p,z_0;z_0)={1\over M_P^3}
\label{b9}\ee
where $M_P$ is the five-dimensional Planck scale in (\ref{e4}).
In this case there is a single multiplicative integration constant left and the standard AdS/CFT procedure extracts from this solution the two-point function of the bulk instanton-density. We denote this bulk axion propagator in the absence of the brane as $G_0(p,z;z_0)$ and satisfies
\be
M_P^3\left[\pa_z^2+\left({Z'\over Z}+4A'\right)\pa_z -e^{-2A}p^2\right]G_0(p,z;z_0)=\delta(z-z_0)
\label{b10}\ee

In our case the presence of an induced action on the SM brane changes the matching conditions to
\be
G_{UV}(p,z_0)=G_{IR}(p,z_0)\sp \partial_zG_{IR}(p,z_0)-\partial_zG_{UV}(p,z_0)={1+(M^2p^2+\Lambda^4)G_{IR}(p,z_0)\over M_P^3}
\label{b11}\ee
The general solution can be written in terms of the bulk propagator $G_0$ with Neumann boundary conditions at the boundary as follows\footnote{Recall that $G(p,z;z_0)$ and $G_0(p,z;z_0)$ are bulk scalar propagators in coordinate space in the radial/holographic direction $z$ and in Fourier space $p^\mu$ for the remaining directions $x^\mu$.}
\cite{CKN}
\be
G(p,z;z_0)={G_0(p,z;z_0)\over 1+(M^2p^2+\Lambda^4)G_{0}(p,z_0;z_0)}
\label{b12}\ee
 The propagator on the brane is obtained by setting $z=z_0$ and becomes
\be
G(p,z_0;z_0)={G_0(p,z_0;z_0)\over 1+(M^2p^2+\Lambda^4)G_{0}(p,z_0;z_0)}
\label{b13}\ee
The general structure of the bulk axion propagator $G_0$ is known, \cite{CKN} and is as follows. The position of the brane $z_0$ determines a bulk curvature energy scale, $R_0$. In the case of simple bulk RG flows\footnote{What is assumed is that the theory does not have multiple intermediate physics scales, but it is controlled by a single mass scale. In the presence of multiple scales a similar analysis is possible.} we obtain
\be
G_0(p,z_0;z_0)={1\over 2 M_P^3}
\left\{ \begin{array}{lll}
\displaystyle {1\over ~p}, &\phantom{aa} &p\gg R_0\\ \\
\displaystyle d_0-d_2p^2-d_4p^4+\cdots,&\phantom{aa}& p\ll R_0.
\end{array}\right.
 \label{b14}
\ee
The IR expansion above is valid for all holographic RG flows. It starts having non-analytic terms starting at $p^4\log p^2$ as is the case with the bulk axion field, \cite{CKN}. The expansion coefficients can be determined either analytically or numerically from the bulk holographic RG flow solution.
The UV expansion in (\ref{b14}) is given, expectantly,  by the flat space result.

 Using (\ref{b14}) we now investigate the axion interaction on the SM brane from (\ref{b13}). It is known that $G_0(p,z_0;z_0)$ is monotonic as a function of $p$, vanishes at large $p$ and attains its maximum at $p=0$ compatible with (\ref{b14}).
Therefore at short enough distances, $p\to\infty$, the axion propagator becomes
\be
G_0(p,z_0;z_0)\simeq {1\over M^2}{1\over p^2}
 \label{b15}\ee
which is the propagator of a massless four-dimensional scalar.
For sufficiently small momenta, $p\ll m$,  we obtain
 \be
G^{-1}(p,z_0;z_0)\simeq  \Lambda^4+2{M_P^3\over d_0}+\left(M^2-2M_P^3{d_2\over d_0^2}\right)p^2+{2M_P^3\over d_0}\left[\left({d_2\over d_0}\right)^2+{d_4\over d_0}\right]p^4+{\cal O}(p^6)
 \label{b16}\ee
 In a simple holographic theory and for the instanton density we have, \cite{CKN}
 \be
 d_0={\bar d_0 \over \ell^3m^4}\sp d_2={\bar d_2 \over \ell^3m^6}
 \sp d_4={\bar d_4 \over \ell^3m^8}
\label{bb19} \ee
 where $m$ is the characteristic scale of the dual QFT, $\ell$ is the IR AdS length, $\bar d_n$ are dimensionless numbers of order ${\cal O}(1)$  and as usual $(M_P\ell)^3\sim N^2$.
The expansion in (\ref{b14}) is valid for $p\ll m$.
We may rewrite (\ref{b16}) using (\ref{bb19}) as
 \be
G^{-1}(p,z_0;z_0)\simeq  \Lambda^4+2{(M_P\ell)^3\over \bar d_0}m^4+\left(M^2-2(M_P\ell)^3{\bar d_2\over \bar d_0^2}m^2\right)p^2+
 \label{bb16}\ee
$$
+{2(M_P\ell)^3\over \bar d_0}\left[\left({\bar d_2\over \bar d_0}\right)^2+{\bar d_4\over \bar d_0}\right]p^4+{\cal O}(p^6)
$$

 We may then recast (\ref{b16}) as the propagator of a massive four-dimensional scalar with effective mass and coupling strengths
\be
f^2_{eff}=M^2+2(M_P\ell)^3{\bar d_2\over \bar d_0^2}m^2\sp m_{eff}^2=
{\Lambda^4+2{(M_P\ell)^3\over \bar d_0}m^4\over f_{eff}^2}\;.
 \label{b17}\ee
 Moreover, the coefficient of the $p^4$ term is dimensionless and of order ${\cal O}(N^2)$.

For $M\gg p\gg m$ we have instead
\be
 G^{-1}(p,z_0;z_0)\simeq \Lambda^4+M^2 p^2+2M_P^3p+\cdots
 \ee

Depending on the hierarchy of the various scales of the problem at intermediate distances, it may be that $(M^2p^2+\Lambda^4)G_{0}(p,z_0;z_0)\ll 1$ and  the axion propagator may behave as a 5-dimensional massless scalar
\be
G_0(p,z_0;z_0)\simeq {1\over 2M_P^3}{1\over p}
 \label{b18}\ee
In such a regime, all axion resonances contribute equitably and the resumed result is as above.

There is a slightly different picture when including the $\eta'$. The proper holographic description that includes both the instanton density and the meson sector, has been discussed in detail in a series of papers, \cite{V-QCD}.
We will not however pursue this further here.

\section{Phenomenological considerations}\label{PhenoConsiderations}

Axions or more generally\footnote{ALP is a two-parameter `model' of (pseudo)scalar particles with very low mass $m_a$ and very weak couplings (suppressed by the mass scale $f_a$) to SM particles.} ALP's can play a significant if not crucial role in cosmology since, depending on their mass and couplings, they can be used to address and solve long-standing problems such as inflation, dark matter and dark energy. In many top-down models (in particular in String Theory or String-derived Supergravity) the number of axions can be significantly larger than one (typically of order $10-10^2$).

In this section we compare our general setup with phenomenological constraints on $f_a$ and $m_a$, \cite{Marsh}. It should be however stressed at this stage, that in many cases this comparison is superficial. The ``kinetic" operators for the axions we consider, can be highly non-standard in some regimes, while almost all experimental constraints have been derived with standard kinetic terms.

The precise role of axions in inflation as well as that in  axion monodromy models, depends on the details of the axion potential, something that  is beyond the scope of the present investigation. Relevant couplings, such as $g_{a\gamma}$ (axion-photon) and $g_{a\ell}$ (axion-lepton), that can be probed in direct detection and/or (local) astrophysical experiments, are  model dependent and this is not something we have analyzed here.

The Weak Gravity Conjecture \cite{swamp-1,swamp-2,swamp-3} states that the decay constant $f_a$ cannot be larger than the Plank scale
\be
f_a < M_P \approx 10^{19} \text{ GeV}
\label{b19}\ee

To address dark energy and ultra-light dark matter, the relevant range for ultra-light axions (ULA) is
\be
 10^{-33} \text{ eV} < m_a < 10^{-18} \text{ eV}
\label{b20}\ee
the lower bound being set by $m_a>M_H$ (from CMB, axion dark energy) and the upper bound from Baryon Jeans scale (from constraints on large scale structure (LSS) formation and the Epoch of Reionization (EoR)). $M_H$ is defined in terms of today's Hubble-Lema\^itre constant to be
\be
H(t_0)= H_0 = 100\, h \text{ (km/s)}/\text{Mpc} = 2.13 \,h \, 10^{-33} \text{ eV} = h M_H
\label{b21}\ee
with a typical value being $h=0.67$ (or lower).

The lower end, $10^{-33} \text{ eV} \,{<}\, m_a \,{<}\, 10^{-30} \text{ eV}$, corresponds to axions that could account for the present Dark Energy (cosmological constant), the decay constant is required to be $f_a \sim M_P$. The region from $m_a\approx 10^{-24} \text{ eV}$ up to $m_a\approx 10^{-18} \text{ eV}$ corresponds to axions that could be good candidates for Dark Matter. A typical upper bound on axion density and the decay constant are $\Omega_a h^2 < 0.12$ and $f_a \leq 10^{16} \text{ GeV}$.

Black hole superradiance is a way to exclude light bosonic fields based only on gravitational interactions. Massive bosonic fields can form bound states around black holes and their mass leads to the existence of stable orbits. Infalling scalars extract energy via the Penrose process. Black hole superradiance for stellar and supermassive black holes respectively excludes the ranges of masses $6 {\cdot}10^{-13} \text{ eV} \,{<}\, m_a \,{<}\, 2 {\cdot} 10^{-11}  \text{ eV}$ and $10^{-18}  \text{ eV} \,{<}\, m_a \,{<}\, 10^{-16}  \text{ eV}$.

\begin{figure}[t]
\centering
\includegraphics[width=0.65\textwidth]{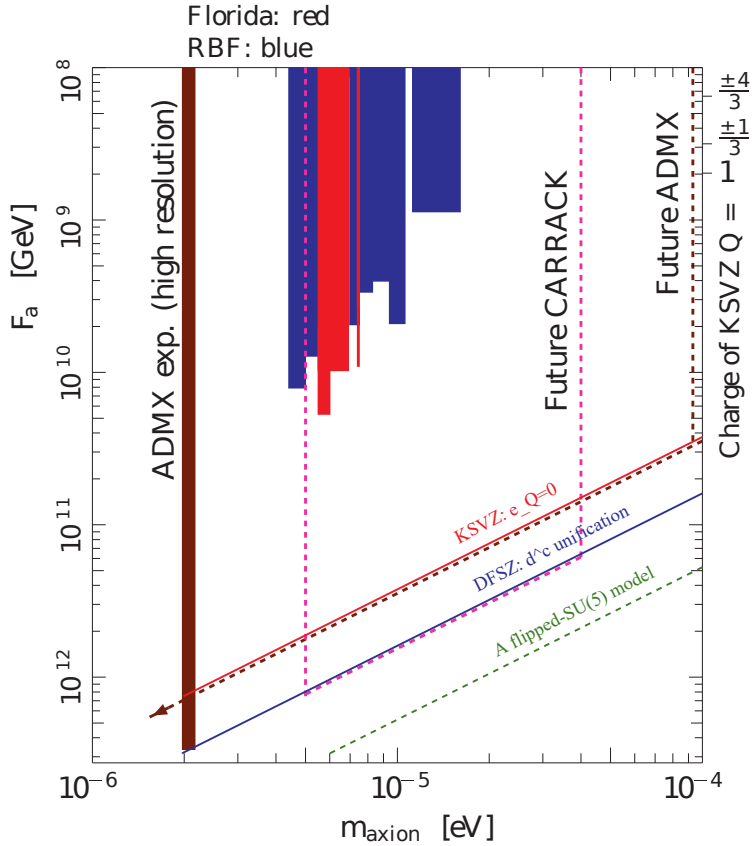}
\vspace{-.1cm}
\caption{Experimental constraints on the axion parameters, $f_a,m_a$. Adapted from \cite{Kim2}.}
\label{fp}
\end{figure}

\paragraph{The QCD axion}

Among the `historical' QCD axion models: PQWW \cite{PQ,Weinberg:1977ma,Wilczek:1977pj} has been  ruled out by experiments, while KSVZ \cite{Kim,SVZ} (heavy quarks and PQ scalar), as well as DFSZ \cite{DFS,Zhitnitsky} (two Higgses and PQ scalar), are still viable, fig \ref{fp}.

In the case where the PQ symmetry is broken ($f_a > H_I/2\pi$) during inflation (which implies free, possibly small initial vacuum misalignment angle\footnote{At the beginning of inflation $\langle\phi_i^2\rangle= f_a \langle\vartheta_{a,i}^2\rangle+ {H_I\over 2\pi}$, where the last term comes from quantum fluctuations of (nearly) massless scalar fields ($m_a\ll H_I$) in De Sitter inflationary spaces.} $\langle\vartheta_{a,i}^2\rangle$, isocurvature perturbations and effects on inflation) one has (no tuning)
\be
10^9  \text{ GeV} < f_a < 10^{15} \text{ GeV}
\label{b22}\ee
On the contrary the standard axion window for unbroken PQ symmetry ($f_a < H_I/2\pi$) during inflation (which implies fixed and large $\langle\vartheta_{a,i}^2\rangle=\pi^2/3$, phase transition relics and axion mini clusters) reads
\be
8{\cdot} 10^9  \text{ GeV} < f_a < 8.5 {\cdot} 10^{10}  \text{ GeV}
\label{b23}\ee
The Anthropic Axion window (that allows favourable conditions for structure formation and life) is instead
\be
f_a < 10^{15}  \text{ GeV} \, .
\label{b24}\ee
Higher values of $f_a$ require $\langle\vartheta_{a,i}^2\rangle$ tuned to be very small.
Constraints from CMB yield
\be
f_a  < 1.4{\cdot} 10^{13}  \text{ GeV}
\label{b25} \ee

We can summarize all the constraints above as
\bea
&&10^{-12}  \text{ eV} < m^{QCD}_a < 10^{-3} \text{ eV}~~~~~
\nn\\
&&10^{9}  \text{ GeV}< f^{QCD}_a < 10^{15}  \text{ GeV}
\label{b27}\eea
 so that $10^{-6}  \text{ GeV}< \Lambda^{QCD}_a < 10^2  \text{ GeV}$, indeed $\Lambda^{QCD}_a =
(m_u \Lambda^3_{QCD})^{1/4}  \approx 10^{-1}  \text{ GeV}$.
The upper bound on the mass here comes from the one-loop anomaly diagram that correlates $f_a$ with $ m^{QCD}_a$. The lower bound for $f^{QCD}_a$ comes from supernovae cooling and the upper bound on $f_a$ from black hole superradiance arguments.

\paragraph{Heavy axions}

There could also exist heavy axions with masses larger than $1 \text{ eV}$, \cite{Millea:2015qra}.
Heavy axions can be unstable on cosmological timescales and decay to lighter particles. The decay increases the relativistic energy density in the Universe and can affect the baryon-photon ratio and baryon abundance. Energy injections due by these decays can deform the shape of the CMB spectrum in such a way that it is not a perfect black body anymore.

The CMB and D/H data combined rule out any energy injection after neutrino decoupling, the allowed masses and lifetimes are the following
\be
m_a> 10 \text{ MeV} \quad \text{ and } \quad \tau_{a \gamma}< 10^{-2} \text{ s}
\label{b28}\ee
where $\tau_{a \gamma}$ is the lifetime for decay into photons.
 It can be computed from the associated decay width as, \cite{rev3},
 \be
 \Gamma_{a\to\g\g}={1\over \tau_{a\g}}={G^2_{a\g\g}m_a^3\over 64\pi}
 \label{b29}\ee
 where $G_{a\g\g}$ is defined as the coupling of the axion to the EM CP-odd density
\be
S_{CP-odd}={G_{a\g\g}\over 4}a~F_{\m\n}\tilde F^{\m\n}
\label{b30}\ee
Typically, it is of order $G_{s\g\g}\sim {\alpha_{\tiny em}\over f_a}$.

 The bound \eqref{b28} assumes no other radiation besides axions and holds until masses become small enough or lifetimes long enough that decays happen after CMB last scattering. In the opposite case, small masses or long lifetimes, the allowed axion masses and lifetimes are
\be
m_a< 10 \text{ eV} \quad \text{ or } \quad \tau_{a \gamma} > 10^{24} \text{ s}
\label{b31}\ee

\subsection{Summary}

There are five `classes' of axion windows for various $m_a$, $f_a$ and $\Lambda_a \equiv \sqrt{m_a f_a}$.

\begin{itemize}
\item \textit{Dark Matter axions (with QCD constraints)}.
\bea
&&10^{-25} \text{ eV} < m^{DM}_a < 10^{-18} \text{ eV} ~~~~~
\nn\\
&&10^{10} \text{ GeV} < f^{DM}_a < 10^{16} \text{ GeV}
\label{b32}\eea
\item \textit{Dark Energy axions (with QCD constraints)}.
\bea
&&10^{-33} \text{ eV} < m^{DE}_a < 10^{-30} \text{ eV} ~~~~~
\nn\\
&&10^{10} \text{ GeV} < f^{DE}_a < 10^{15} \text{ GeV}
\label{b34}\eea

\item \textit{Axions as Inflatons}. These are highly model dependent and there are no strict bounds.
\item \textit{Heavy Axions}. There could also exist heavy axions with masses larger than $ 1 \text{ eV}$. The allowed axion masses and lifetimes are
\be
m_a> 10 \text{ MeV} \quad \text{ and } \quad \tau_{a \gamma}< 10^{-2} \text{ s}
\label{heavy_cond_1}
\ee
or
\be
m_a< 10 \text{ eV} \quad \text{ or } \quad \tau_{a \gamma} > 10^{24} \text{ s}
\label{b35}\ee

\item \textit{QCD axion}.
\bea
&&10^{-12} \text{ eV} < m^{QCD}_a < 10^{-3} \text{ eV} ~~~~~
\nn\\
&&10^{9} \text{ GeV} < f^{QCD}_a < 10^{15}\text{ GeV}
\label{b36}\eea

\end{itemize}

\subsection{Comparison with the generic composite axions}

We now contrast our previous frameworks with the various phenomenological constraints.

\paragraph{Composite Axions.}

For composite generic axions with $m_h$, $ m_{_{SM}} \ll M$ we have from \eqref{a28}
\be
m_a^2 = {\bar{a}_0\over  \bar{a}_2} m_h^2 \quad , \quad f^2_a = {\bar{a}_2\over  \bar{a}_0} {m_h^2\over \lambda_0^2} \left({M\over m_h}\right)^{2\Delta_h}
\label{b37}\ee
 Assuming ${\bar{a}_0/\bar{a}_2}, \lambda_0\approx 1$, and $\Delta_h\approx 4$, the mass $m_a$ is completely determined by the hidden mass scale $m_h$ and the decay constant $f_a$ can be hierarchically larger thanks to the enhancement factor $(M/m_h)^4$.  The messenger scale is given by
\be
M\approx (f_a m_h^3)^{1/4} \approx (f_a m_a^3)^{1/4}
\label{messangers_mass}
\ee
From the assumption $ m_{_{SM}} \ll M$ and the weakest upper bound $f_a < M_P$, we obtain a general lower bound for the axion mass
\be
m_a \gg 10 \text{ eV}
\label{composite_mass_bound}
\ee
Therefore composite axions can be generically compared to heavy axions scenarios only.
However,  the CMB case for heavy axions \eqref{heavy_cond_1} is allowed provided the bound on $\tau_{a\gamma}$ is small enough. This can be achieved if $m_a$ is sufficiently high, or there is a symmetry reason for suppressing the axion coupling to two photons.

In the case $m_h<m_{_{SM}}$, considering again $f_a < M_P$, the messenger scale has an upper bound
\be
M \lesssim 10 \text{ TeV}
\label{b38}\ee
which satisfies also $m_{_{SM}}\ll M$. In the case $m_h>m_{_{SM}}$ the assumptions are satisfied taking $f_a \gg m_a$. $M$, $m_a$ and $f_a$ can range from $m_{_{SM}}$ to the Plank scale.

For the QCD axion the messenger scale can be written as $M^2\sim m_a \Lambda_a$, where
\be
\Lambda_a=\sqrt{m_a f_a}\,{\sim}\,(m_u \Lambda^3_{QCD})^{1/4} \,{\sim}\, 10^{-1} \text{ GeV} \,{\sim}\, m_{_{SM}}
\label{b39}\ee
In this case the conditions $m_h$, $ m_{_{SM}} \ll M$ cannot be satisfied together.

An obvious issue here is the assumption $p \ll m_h$ in the formulae used. It is not clear whether this holds once for example when $m_h \sim 10^{- 23} \text{ eV}$. If we instead use the formulae $m_h \ll p \ll m_{_{SM}}$, the axion is unfortunately non-local and we need a more careful first principles computation to determine the viability of our model.

\paragraph{Holographic Axions.}

For holographic axions with $p < m_h  \ll M$ we have from \eqref{b17}
\be
m_a^2 f_a^2 = \Lambda^4 + {2\over \bar{d}_0} (M_P\ell)^3 m_h^4  \quad , \quad f_a^2 = M^2 + {2 \bar{d}_2\over \bar{d}_0^2}(M_P\ell)^3 m_h^2
\label{b40}\ee
where $\Lambda,M$ are parameters of the brane (visible) theory while $(M_P\ell)^3,m_h$ are parameters of the bulk (hidden) theory.

Eliminating the combination $M_P\ell$ from these equations, the messenger scale turns out to be
\be
M = \sqrt{ f_a^2\left(1 - \gamma {m_a^2\over m_h^2}\right) + \gamma {\Lambda^4\over m_h^2}}  =
\sqrt{ f_a^2 + \gamma {\Lambda^4-\Lambda_a^4 \over m_h^2}}
\label{b41}\ee
where $\gamma=\bar{d}_2/\bar{d}_0$ is a constant of order 1 and $\Lambda_a=\sqrt{m_a f_a}$.  Assuming $\Lambda^4 = \kappa \Lambda^4_a$ with $\kappa$ of order 1, one eventually finds
\be
M = f_a\sqrt{ 1 + \gamma\kappa {m_a^2\over m_h^2}}
\label{bb42}\ee
that produces in all cases (QCD, DM and DE) reasonable messenger scales $M\approx f_a$ for any $m_a \le m_h$. For $m_a \gg m_h$ we still obtain a reasonable relation $M \approx \Lambda_a^2/m_h = f_a (m_a /m_h) \gg f_a$.

For $ m_h < p < M$, we obtain  an effective 5-dimensional axion.

\vskip .2cm
\section*{\bf Acknowledgments}
\addcontentsline{toc}{section}{Acknowledgments}

We would like to thank, George Anastasopoulos, Asimina Arvanitaki, Matteo Baggioli, Paolo~Di~Vecchia, Jewel Kumar Ghosh, Yuta Hamada, Vassilis Niarchos, Francesco Nitti, Maxim Pospelov, Giancarlo Rossi, Alberto Salvio, Misha Shifman, Arcady Vainshtein, Gabriele Veneziano, Giovanni Villadoro and Lukas Witkowski for useful discussions.
We would also like  to thank the anonymous referee, whose remarks and criticism helped us improve the presentation in this paper.

This work was supported in part by  the Advanced ERC grant SM-grav, No 669288. P.A. and D.C. was supported by FWF Austrian Science Fund via the SAP P30531-N27.

M.B. and D.C. would like to thank the MIUR-PRIN contract 2015MP2CX4002 {\it ``Non-perturbative aspects of gauge theories and strings''} for partial support.

\noindent

 \newpage
\appendix

\renewcommand{\theequation}{\thesection.\arabic{equation}}
\addcontentsline{toc}{section}{Appendices}

\section*{APPENDIX}

\section{Cancellation of gauge anomalies in the messenger sector\label{anomalies}}

In this appendix we discuss the detailed anomaly cancelation requirements for the messenger sector and its coupling to the SM.

We assume that QFT$_N$ and the SM are separately anomaly free. We must impose that the messenger fermions do not introduce gauge anomalies.
For the QFT$_N$ gauge group (that we assume to be SU(N)) the condition is that there is an equal number of $N$ and $\bar N$.
Similarly, anomaly freedom for the SM implies that there must be an equal number of $3$ and $\bar 3$, $2$ and $\bar 2$ (for the cancelation of the Witten SU(2) anomaly) and hypercharge $Q$ and $\bar Q$.

\begin{center}
\begin{tabular}[t]{|c|c|c|}
 \trule
 Representation & Number & Charge\\
 \trule
 $(N,3)$& $n_3$&\\
 \trule
$(\bar N,3)$& $\bar n_3$&\\
 \trule
 $(N,\bar 3)$& $n_{\bar 3}$&\\
 \trule
 $(\bar N,\bar 3)$& $\bar n_{\bar 3}$&\\
 \trule
$(N,2)$& $n_2$&\\
 \trule
$(\bar N,2)$& $\bar n_2$&\\
 \trule
 $(N,\bar 2)$& $n_{\bar 2}$&\\
 \trule
 $(\bar N,\bar 2)$& $\bar n_{\bar 2}$&\\
 \trule
 $(N,1)$& $n_1$& $Q_1$\\
 \trule
$(\bar N,1)$& $\bar n_1$& $\bar Q_1$\\
 \trule
\end{tabular}
\label{t1} \end{center}
All spinors are assumed to be left-handed Weyl spinors.
In order to allow masses for all the messengers we must have $n_3=\bar n_{\bar 3}$, $n_{\bar 3}=\bar n_3$, $n_1=\bar n_1$, $Q_1=-\bar Q_1$.

 In terms of the above table of charges the conditions for absence of anomalies become
 \begin{itemize}
 \item  $SU(N)^3$
 \be
3 n_3+ 3 n_{\bar 3}+ 2 n_2+ 2 n_{\bar 2}+ n_1= 3 \bar n_3+ 3 \bar n_{\bar 3}+ 2 \bar n_2+ 2 \bar n_{\bar 2}+\bar n_1\to n_2+n_{\bar 2}=\bar n_2+\bar n_{\bar 2}
\label{1} \ee
\item $SU(3)^3$
 \be
  n_3+\bar n_{ 3}=n_{\bar 3}+\bar n_{\bar  3}~~~{\rm  (automatic)}
 \label{2}\ee
 \item Witten SU(2)
 \be
  n_2+\bar n_{ 2}+n_{\bar 2}+\bar n_{\bar  2}={\rm even}~~~({\rm automatic})
 \label{3}\ee
  \item $U(1)^3$
 \be
  n_1~Q_1^3+\bar n_1 \bar Q_1^3 =0
 \label{4}\ee
 \item $U(1)$ and $SU(N)^2\times U(1)$
  \be
  n_1~Q_1+\bar n_1 \bar Q_1 =0
 \label{5}\ee
 All other mixed anomalies vanish
 \end{itemize}
From (\ref{4}) and (\ref{5}) we obtain
\be
\bar n_1=n_1\sp \bar Q_1=-Q_1
\label{6}\ee

Taking into account anomaly cancellation, we obtain for the following charge assignments
\begin{center}
\begin{tabular}[t]{|c|c|c|}
 \trule
 Representation & Number & Charge\\
 \trule
 $(N,3)$& $n_3$&\\
 \trule
$(\bar N,3)$& $\bar n_3$&\\
 \trule
 $(N,\bar 3)$& $\bar n_{ 3}$&\\
 \trule
 $(\bar N,\bar 3)$& $n_{ 3}$&\\
 \trule
$(N,2)$& $n_2$&\\
 \trule
$(\bar N,2)$& $\bar n_2$&\\
 \trule
 $(N,\bar 2)$& $n_{\bar 2}$&\\
 \trule
 $(\bar N,\bar 2)$& $n_2+n_{\bar 2}-\bar n_2$&\\
 \trule
 $(N,1)$& $n_1$& $Q$\\
 \trule
$(\bar N,1)$& $ n_1$& $- Q$\\
 \trule
\end{tabular}
\label{t2} \end{center}

If we include a ``realistic" SM sector that is fully bifundamental, then we know from previous work on orientifold embeddings of the SM that it must include at least a couple of extra ``anomalous" U(1)s, \cite{pa,class,KA}. These give rise to more anomalies to be considered. We shall not consider them further in this paper.

Ignoring the SM interactions (that are weak) the messengers have a fermionic chiral symmetry that may be  broken by mass terms to a vectorial subgroup.

The most important  point here is that the overall  $U(1)_A$ is anomalous.
We have
\be
\partial_{\mu}J^{\m}=2( n_3+n_{\bar 3}+ n_2+n_{\bar 2}+ 2 n_1)Tr[G\wedge G]+
\label{7}\ee
$$+2N( n_3+\bar n_{ 3})Tr[F_3\wedge F_3]+2N( n_2+
n_{\bar 2})Tr[F_2\wedge F_2]+
$$
$$+2 N n_1Q^2~Tr[F_1\wedge F_1]+{\rm mass~corrections}
$$
This analysis must be amended by replacing the SM with one of the quivers  found in orientifolds, \cite{pa}. There might be new constraints on the extra anomalous U(1)'s that appear in that case.

\section{The calculation of the box diagrams}\label{boxAppendix}

\begin{figure}[h]
\centering
\includegraphics[width=0.45\textwidth]{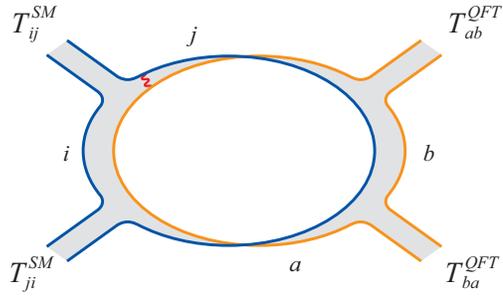}
\vspace{-.1cm}
\caption{The 1-loop diagram with two gauge fields from QFT$_{N}$ and two from the SM, in a ``double line" notation. The $a,b$ and $i,j$ indexes denote the colours of the QFT$_{N}$ and the SM respectively. The lines denote the index contractions of the hidden theory and the SM.}
\label{f0planar}
\end{figure}

In this appendix we present the computation of the effective couplings \eqref{final}. To extract these coupling we compute the scattering amplitude of two gauge fields from QFT$_{N}$ and two from the SM in the low energy approximation (fig \ref{f0planar}).

\begin{figure}[h]
\centering
\includegraphics[width=0.4\textwidth]{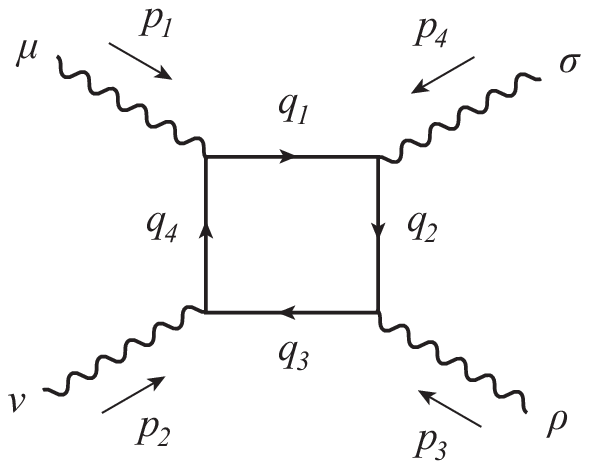}\\
\vspace{.1cm}
\includegraphics[width=0.35\textwidth]{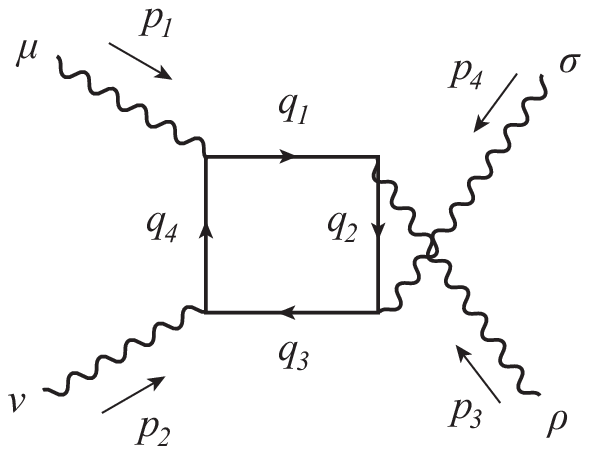}~~~~~~~~~~~~~~~~~~
\includegraphics[width=0.35\textwidth]{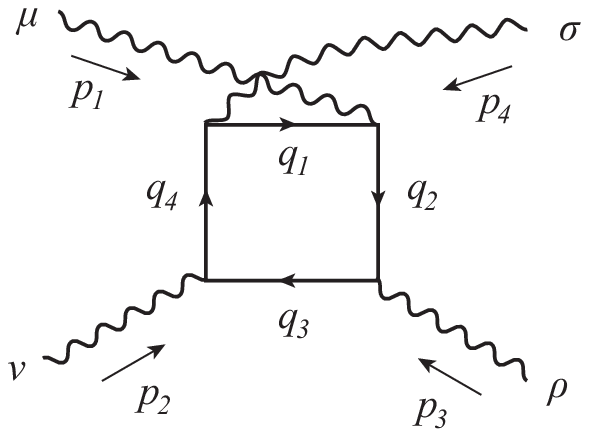}
\caption{ The three box diagrams. Apart from the first ``box-1", we have to add the contribution from two more diagrams with $p_3,\r \leftrightarrow p_4, \s$ (``box-2") and $p_1,\m \leftrightarrow p_4, \s$ (``box-3") exchanged.}
\label{f1}
\end{figure}

The leading terms are six box diagrams, three of them are represented in fig \ref{f1}, the other three diagrams are similar to the previous ones but with the internal fermion line going anti-clockwise instead of clockwise.
The full amplitude can be written as
\begin{align}
A=\, &  \tr (T_1^{\textup{SM}} T_2^{\textup{SM}} T_3^\textup{QFT} T_4^\textup{QFT}) A^{(1234)}+
\tr (T_1^{\textup{SM}} T_2^{\textup{SM}} T_4^\textup{QFT} T_3^\textup{QFT}) A^{(1243)}+\nn \\
&\tr (T_1^{\textup{SM}} T_3^\textup{QFT} T_2^{\textup{SM}} T_4^\textup{QFT}) A^{(1324)}+
\tr (T_1^{\textup{SM}} T_3^\textup{QFT} T_4^\textup{QFT} T_2^{\textup{SM}}) A^{(1342)}+\nn \\
&\tr (T_1^{\textup{SM}} T_4^\textup{QFT} T_2^{\textup{SM}} T_3^\textup{QFT}) A^{(1423)}+
\tr (T_1^{\textup{SM}} T_4^\textup{QFT} T_3^\textup{QFT}  T_2^{\textup{SM}}) A^{(1432)}
\label{allboxtogether}
\end{align}
We can hold $T_1^{\textup{SM}}$ in the first position using cyclicity of the trace. However all the traces yield the same result because they split in two independent traces, for instance
\begin{equation}
\tr (T_1^{\textup{SM}} T_2^{\textup{SM}} T_3^\textup{QFT} T_4^\textup{QFT})=
\tr (T_1^{\textup{SM}} T_2^{\textup{SM}})
\tr (T_3^\textup{QFT} T_4^\textup{QFT}) \label{TraceSplit}
\end{equation}
Up to traces,  there is no difference between the same diagrams computed with two pairs of different non-abelian bosons or four identical abelian bosons. The result holds even for the non-abelian case\footnote{Of course, in the non-abelian case there will be contributions from higher diagrams covariantizing the result.}, since the gauge groups live on different sectors, and split to separate traces like in \eqref{TraceSplit}. Focusing on the first color-ordered diagram $A^{(1234)}$ (box-1 in figure \ref{f1}) we have
\bea
i A^{(1234)} {=}
 {-} \text{Tr}\left[\eslash_1{1{+}\gamma^5\over 2}{i\over \qslash_1 {+}M}  \eslash_4{1{+}\gamma^5\over 2}{i\over \qslash_2 {+}M} \eslash_3{1{+}\gamma^5\over 2}{i\over \qslash_3 {+}M}
\eslash_2{1{+}\gamma^5\over 2}{i\over \qslash_4 {+}M}\right]
\label{BoxAmplitude}
\eea
with
\be
q_1=p\sp q_2=p+p_4\sp q_3=p-p_1-p_2 \sp q_4=p_1-p \label{internalmomenta}
\ee

The amplitude is a sum of several terms with different number of $\gamma_5$ in the trace. After some computations we get two different classes of terms, with no (scalar) or one $\gamma_5$ (proportional to $\e_{\m\n\r\s}$). In the non-chiral case, the second part vanishes.

Following the usual procedure we perform the Feynman trick by shifting the loop momentum $p\to P$ in order to have perfect squares of $P$ in the denominator, and we drop all odd parts of $P$ which vanish upon integration. The amplitude becomes
\bea
A^{(1234)}&=& 3! i \mu^{4-D} \int_{[0,1]^4} dx_1 \dots dx_4 \int \frac{d^D P}{(2\pi)^D} \frac{\AA_{\mu \nu \rho \sigma} P^\mu P^\nu P^\rho P^\sigma+
\BB_{\mu \nu } P^\mu P^\nu +
\CC}{(P^2-\widetilde{M}^2)^4}~~~~~~~~~~
\\
&=& -\frac{1}{(4\pi)^2} \int_{[0,1]^4} dx_1 \dots dx_4 \times \label{A1234afterPintegration}\\
&& \Bigg[
{\AA_{\mu \nu \rho \sigma}\frac{g^{\mu\nu}g^{\rho\sigma}+g^{\mu\rho}g^{\nu\sigma}+g^{\mu\sigma}g^{\nu\rho}}{4}}
\left(\frac{1}{\epsilon} +\log \frac{4\pi \mu^2 e^{ -\gamma_E}}{\widetilde M^2}\right){-}\frac{g^{\mu\nu}{\BB}_{\mu\nu}}{\widetilde M^2}{+}\frac{{\CC}}{\widetilde M^4}\Bigg]\nn
\eea
where $\widetilde M^2=M^2-2x_3(1-x_3-x_4)p_1{\cdot}p_2 - 2x_2 (x_3+x_4) p_1{\cdot}p_4 - 2 x_2 x_3 p_2{\cdot}p_4$.
$\AA_{\mu \nu \rho \sigma}$, $\BB_{\mu \nu}$ and $\CC$ are functions of the external momenta, the external polarizations and the Feynman parameters. For dimensional reasons $\AA$ does not contain $M$ while $\BB$ and $\CC$ have at most $M^2$ and $M^4$ terms respectively.
The integral over the momenta of the term $P^\mu P^\nu P^\rho P^\sigma$ is divergent thus we use dimensional regularization to integrate over the loop momentum and get \eqref{A1234afterPintegration}.
The divergence cancels when we sum all box diagrams in figure \ref{f1}.

In the low energy limit we are in the regime $M^2\gg p^2$. The expansion our expressions brings $\Delta$'s in the numerator. Therefore, all terms in \eqref{A1234afterPintegration} become polynomials of the Feynman parameters $x_1,x_2,x_3,x_4$ which can be easily integrated out giving the final result for the amplitude
\begin{equation}
A^{(1234)}=A^{(1234)}_{{{(0)}}} + \frac{A^{(1234)}_{{{(2)}}}}{ M^2} + \frac{A^{(1234)}_{{{(4)}}}}{ M^4}+O(M^{-6})
\label{A1234final}
\end{equation}
Adding all box diagrams we get the final result which has the same form as \eqref{A1234final}.
\begin{equation}
A=\tr(T_1^{\textup{SM}} T_2^{\textup{SM}})
\tr(T_3^\textup{QFT} T_4^\textup{QFT})
\left( A_{{(0)}} + \frac{A_{{(2)}}} {{M^2}} + \frac{A_{{(4)}}}{ M^4}+O(M^{-6}) \right)
\label{AfinalAppendix}
\end{equation}
$A_{{(0)}}$ and $A_{{(2)}}$ vanish due to gauge invariance. The explicit expression of $A_{{(4)}}$ is complicated, in order to extract the effective coupling  we use the spinor-helicity formalism \cite{spinor_helicity, Bianchi:2008pu, Bianchi:2015vsa}. It helps to expose a great number of cancellations that remove superfluous terms thanks to smart kinematic/gauge choices.

\subsection{The box amplitude in different helicity configurations}
\label{box_helicity}

In this section we evaluate the amplitude in different helicity configurations. Our results assume very compact and neat forms which can be used to evaluate various couplings in the effective action.

In principle, one has to deal with 16 different helicity configurations $({\pm}{\pm}{\pm}{\pm})$. However, those with the same number of $+$'s and $-$'s are connected by permutations. In addition, configurations which are related via parity $+\leftrightarrow -$ yield similar results. Therefore, we end up with three different classes of configurations, with four/zero, three/one and two/two $+$'s and $-$'s.

\paragraph{Helicity configurations $({+}{+}{+}{+})$ and $({-}{-}{-}{-})$.}

When all helicities have the same sign, the `gauge' choice of auxiliary momenta $q_i{=}q {\neq} p_i$ for all $i=1,...4$ makes all the scalar products between polarizations vanish
\begin{equation}
\e_+(p_i){\cdot}\e_+(p_j)=0
\end{equation}
In principle we have 81 terms of the form $\e_1{\cdot} p_i \e_2{\cdot} p_j \e_3{\cdot} p_k \e_4{\cdot} p_l$. Among them there are combinations that are independent on the auxiliary momentum $q_i{=}q$ in the polarizations. In these, all momenta are different, for example
\bea
\e_1^{(+)}{\cdot} p_4 \e_2^{(+)}{\cdot} p_3 \e_3^{(+)}{\cdot} p_2 \e_4^{(+)}{\cdot} p_1&=& \frac{1}{4} \dfrac{\<q p_4\>[p_4 p_1]}{\<q p_1\>}
\dfrac{\<q p_3\>[p_3 p_2]}{\<q p_2\>}
\dfrac{\<q p_2\>[p_2 p_3]}{\<q p_3\>}
\dfrac{\<q p_1\>[p_1 p_4]}{\<q p_4\>}\nn\\
&=&\frac{1}{4} [p_1 p_4]^2[p_2 p_3]^2
\eea
where $\<q p\>$ and $[pq]$ are the invariant spinorial contractions. Using momentum conservation we can express all the terms like $\e_1{\cdot} p_i \e_2{\cdot} p_j \e_3{\cdot} p_k \e_4{\cdot} p_l$ as a linear combination of the following nine terms
\begin{align}
\e_1{\cdot} p_4 \e_2{\cdot} p_3 \e_3{\cdot} p_2 \e_4{\cdot} p_1 \qquad
\e_1{\cdot} p_4 \e_2{\cdot} p_3 \e_3{\cdot} p_1 \e_4{\cdot} p_2 \qquad
\e_1{\cdot} p_4 \e_2{\cdot} p_1 \e_3{\cdot} p_2 \e_4{\cdot} p_3\\
\e_1{\cdot} p_3 \e_2{\cdot} p_4 \e_3{\cdot} p_1 \e_4{\cdot} p_2 \qquad
\e_1{\cdot} p_3 \e_2{\cdot} p_1 \e_3{\cdot} p_4 \e_4{\cdot} p_2 \qquad
\e_1{\cdot} p_3 \e_2{\cdot} p_4 \e_3{\cdot} p_3 \e_4{\cdot} p_1\\
\e_1{\cdot} p_2 \e_2{\cdot} p_3 \e_3{\cdot} p_4 \e_4{\cdot} p_1 \qquad
\e_1{\cdot} p_2 \e_2{\cdot} p_1 \e_3{\cdot} p_4 \e_4{\cdot} p_3 \qquad
\e_1{\cdot} p_2 \e_2{\cdot} p_4 \e_3{\cdot} p_1 \e_4{\cdot} p_3
\end{align}
which, however, are not independent. Using
\begin{itemize}
\item momentum conservation
\item anti-symmetry of spinor product $[pq]=-[qp]$, \footnote{For instance
\bea
\e_1{\cdot} p_4 \e_2{\cdot} p_3 \e_3{\cdot} p_1 \e_4{\cdot} p_2&=& \frac{1}{4}[p_1 p_4][p_2 p_3][p_3 p_1][p_4 p_2]\nn\\
&=& \frac{(-1)^4}{4}[p_4 p_1][p_3 p_2][p_1 p_3][p_2 p_4]=
\e_1{\cdot} p_3 \e_2{\cdot} p_4 \e_3{\cdot} p_2 \e_4{\cdot} p_1 ~~~~~
\eea}
and
\item Schouten identity $[p_1 p_2][p_3 p_4]+[p_2 p_3][p_1 p_4]+[p_3 p_1][p_2 p_4]=0$
\end{itemize}
we can extract three independent combinations
\begin{align}
%
\e_1{\cdot} p_2 \e_2{\cdot} p_1 \e_3{\cdot} p_4 \e_4{\cdot} p_3&=\frac{1}{4}[p_1 p_2]^2[p_3 p_4]^2 \\
\e_1{\cdot} p_3 \e_2{\cdot} p_4 \e_3{\cdot} p_2 \e_4{\cdot} p_3&=\frac{1}{4}[p_1 p_3]^2[p_2 p_4]^2 \\
\e_1{\cdot} p_4 \e_2{\cdot} p_3 \e_3{\cdot} p_2 \e_4{\cdot} p_1&=\frac{1}{4}[p_1 p_4]^2[p_2 p_3]^2
\end{align}
%
%
%
After all these manipulations the amplitude reads
\bea
&& A_{{{(0)}}}^{++++}=A_{{(2)}}^{++++}=0 \\
&& A_{{{(4)}}}^{++++}= -\frac{1}{30} \Big([p_1 p_2]^2 [p_3 p_4]^2+[p_1 p_3]^2 [p_2 p_4]^2+[p_1 p_4]^2 [p_2 p_3]^2\Big)
\eea
By parity/conjugation $[p_ip_j]\rightarrow\<p_i p_j\>$, the all-minus configuration produces a similar result
\bea
&& A_{{{(0)}}}^{----}=A_{{{(2)}}}^{----}=0 \\
&& A_{{{(4)}}}^{----}= -\frac{1}{30} \Big(\<p_1 p_2\>^2 \<p_3 p_4\>^2+\<p_1 p_3\>^2 \<p_2 p_4\>^2+\<p_1 p_4\>^2 \<p_2 p_3\>^2\Big)
\eea

\paragraph{Helicity configurations $({-}{+}{+}{+})$ and $({+}{-}{-}{-})$.}

In principle one we have 8 configurations, however not all are independent because they are connected by cyclic transformations. The only two independent configurations are $({-}{+}{+}{+})$ and $({+}{-}{-}{-})$.

In both cases we choose $q_1{=}p_3$ and $q_2{=}q_3{=}q_4{=}p_1$ as auxiliary momenta, which leads to
\begin{equation}
\e(p_i){\cdot}\e(p_j)=0 \qquad \e(p_1){\cdot}p_3=0 \qquad \e(p_i){\cdot}p_1=0
\end{equation}
Evaluating the amplitudes we obtain
\bea
&& A^{-+++}_{{{(0)}}}=A^{-+++}_{{{(2)}}}=A^{-+++}_{{{(4)}}}=0\\
&& A^{+---}_{{{(0)}}}=A^{+---}_{{{(2)}}}=A^{+---}_{{{(4)}}}=0
\eea
as expected.

\paragraph{Helicity configuration $({-}{-}{+}{+})$.}

The four helicity configurations with two consecutive pluses and minuses can be obtained by cyclic permutations of  $({-}{-}{+}{+})$.

In this case we choose $q_1{=}q_2{=}p_3$ and $q_3{=}q_4{=}p_2$ as auxiliary momenta so that all the products between polarizations are zero except $\e_+(p_1){\cdot}\e_-(p_4)$
\bea
\e(p_i){\cdot}\e(p_j)=0 \,\, (i{\neq}1,j{\neq}4) &\qquad& \e_+(p_1){\cdot}p_3=0 \qquad \e_+(p_2){\cdot}p_3=0 \nn\\
&\qquad& \e_-(p_3){\cdot}p_2=0 \qquad\e_-(p_4){\cdot}p_2=0
\eea
With these choices we obtain
\bea
&&A_{{{(0)}}}^{--++}=A_{{{(2)}}}^{--++}=0 \\
&& A_{{{(4)}}}^{--++}=-\frac{2}{5} p_1{\cdot} \e_2 p_2{\cdot} \e_1 p_3{\cdot} \e_4 p_4{\cdot} \e_3+\frac{17}{45} t p_1{\cdot} \e_2 p_4{\cdot} \e_3 \e_1{\cdot} \e_4
\eea
The scalar products in the last formula are not independent, indeed
\begin{align}
p_1{\cdot} \e_2 p_2{\cdot} \e_1 p_3{\cdot} \e_4 p_4{\cdot} \e_3=& \frac{1}{4} \<p_1 p_2\>^2 [p_3 p_4]^2=
\frac{t}{4} p_1{\cdot} \e_2 p_4{\cdot} \e_3 \e_1{\cdot} \e_4
\end{align}
where $t = -2p_2p_3 = - \<p_2 p_3\> [p_2 p_3] = -2p_1p_4 = - \<p_1 p_4\> [p_1 p_4]$. We finally get
\bea
&& A_{{{(0)}}}^{--++}=A_{{{(2)}}}^{--++}=0 \\
&& A_{{{(4)}}}^{--++}=\frac{5}{18} \<p_1 p_2\>^2 [p_3 p_4]^2
\eea
and similarly for $({+}{+}{-}{-})$
\bea
&& A_{{{(0)}}}^{++--}=A_{{{(2)}}}^{++--}=0 \\
&& A_{{{(4)}}}^{++--}=\frac{5}{18}  [p_1 p_2]^2 \<p_3 p_4\>^2
\eea

\paragraph{Helicity configuration $({-}{+}{-}{+})$.}

We have two configurations with alternating positive and negative helicities $({+}{-}{+}{-})$ and $({-}{+}{-}{+})$ which are related by cyclic permutations.

In the configuration $({-}{+}{-}{+})$ we choose $q_1{=}q_3{=}p_2$ and $q_2{=}q_4{=}p_3$ as auxiliary momenta obtaining
\bea
\e(p_i){\cdot}\e(p_j)=0 \quad (i{\neq}1,j{\neq}4) &\qquad& \e_+(p_1){\cdot}p_2=0 \qquad\e_-(p_2){\cdot}p_3=0\nn\\
&\qquad& \e_+(p_3){\cdot}p_2=0 \qquad\e_-(p_4){\cdot}p_3=0
\eea
The amplitude reads
\bea
A_{{{(0)}}}^{-+-+}&=&A_{{{(2)}}}^{-+-+}=0 \\
A_{{{(4)}}}^{-+-+}&=&-\frac{2}{5} p_1{\cdot} \e_2 p_4{\cdot} \e_1 p_1{\cdot} \e_4 p_4{\cdot} \e_3+\frac{17}{45} t p_1{\cdot} \e_2 p_4{\cdot} \e_3 \e_1{\cdot} \e_4
=\frac{5}{18}  \<p_1 p_3\>^2 [p_2 p_4]^2 ~~~~
\eea
which is related to the $({-}{-}{+}{+})$ amplitude by the exchange $p_2\leftrightarrow p_3$.

\subsection{Effective action}

In this section we compute the effective action after integrating out the heavy fermions (messengers). Our goal is to extract the couplings in the action by comparing them with the box amplitudes evaluated above in \ref{box_helicity}.

Integrating out the heavy fermions, the effective Lagrangian takes the form
\begin{align}
\LL_\textup{eff}=
\frac{g_\textup{SM}^2 g_\textup{QFT}^2}{M^4}\Big[& a_1 (F \,F)(G\, G)+a_2 (F\, G)^2+  i b_1 (F\,\widetilde{F})(G \,G)+i b_2 (F \,F) (G\,\widetilde{G})+\nonumber \\
&+i b_3 (F \,G)(F\,\widetilde{G})+ c_1 (F\,\widetilde{F})(G\,\widetilde{G})+c_2 (F\,\widetilde{G})^2\Big]
\label{11}\end{align}
where $ (F\,G)=F_{\mu \nu} G^{\mu \nu}$, $\widetilde{F}_{\mu\nu}={1\over 2}\epsilon_{\mu\nu\rho\sigma}F^{\rho\sigma}$. $a_i$, $b_i$ and $c_i$ are real coefficients. In the above $F_{\m\n}$ are the (abelianized) SM field strengths, while $G_{\m\n}$ are the (abelianized) field strengths of QFT$_N$.

Each field strength can be split into self-dual and anti-self-dual components, $F=F_+{+}F_-$, where $F_\pm = {1\over 2}(F \pm i \widetilde{F})$ are defined according to
\begin{equation}
(\widetilde{F}_\pm)_{\mu \nu}= \frac{1}{2} \epsilon_{\mu \nu \rho \sigma} F_\pm^{\rho \sigma}= \pm i (F_\pm)_{\mu \nu}
\end{equation}
When the field strength is associated to massless vector field we can identify helicity with chirality using Weyl spinors.

Using as a basis the self-dual and anti-self-dual Lorentz generators $\sigma^{\mu \nu}$ and $\bar{\sigma}^{\mu \nu}$
in the spinorial representations
\begin{equation}
\frac{1}{2} \epsilon^{\mu \nu \rho \sigma} (\sigma_{\rho \sigma})^\alpha_{\,\,\,\,\beta}=
i (\sigma^{\mu \nu})^\alpha_{\,\,\,\,\beta} \qquad
\frac{1}{2 } \epsilon^{\mu \nu \rho \sigma} (\bar{\sigma}_{\rho \sigma})_{\dot{\alpha}}^{\,\,\,\,\dot{\beta}}=
-i (\bar{\sigma}^{\mu \nu})_{\dot{\alpha}}^{\,\,\,\,\dot{\beta}}
\end{equation}
we can write the field strength as
\begin{equation}
F^{\mu \nu}=
(f_+)^{\alpha \beta} (\sigma^{\mu \nu})_{\alpha \beta}+
(f_-)_{\dot{\alpha} \dot{\beta}} (\bar{\sigma}^{\mu \nu})^{\dot{\alpha} \dot{\beta}}
\end{equation}
Since $(\sigma^{\mu \nu})_{\alpha \beta}$ is symmetric in the spinorial indices, $f_+$ can be thought of as the symmetric tensor product of two left handed commuting Weyl spinors
\be
(f_+)_{\alpha \beta}(k)
= u_\alpha(k) u_\beta(k) = |k]_\alpha |k]_\beta,
 \ee
where $u_\alpha(k)$ is a massless commuting Weyl spinor of definite (positive) helicity $h=+{1\over 2}$ and light-like momentum $k^\mu = u^\alpha(k)\sigma^\mu_{\alpha\dot\alpha} \bar{u}^{\dot\alpha}(k)$, so that $(f_+)_{\alpha \beta}$ has helicity $h=+1$.

Due to Lorentz invariance, tensors with positive and negative helicities (or more correctly left and right chirality) are orthogonal to each other
\begin{equation}
F_{\mu \nu} G^{\mu \nu}= (f_+)^{\alpha \beta} (g_+)^{\gamma \delta} (\sigma^{\mu \nu})_{\alpha \beta} (\sigma_{\mu \nu})_{\gamma \delta}+
(f_-)^{\dot{\alpha} \dot{\beta}} (g_-)^{\dot{\gamma} \dot{\delta}} (\bar{\sigma}^{\mu \nu})_{\dot{\alpha} \dot{\beta}} (\bar{\sigma}_{\mu \nu})_{\dot{\gamma} \dot{\delta}}
\end{equation}
therefore, the terms that appear in the effective action (\ref{11}) read
\begin{align}
(F\,G)=&F_{\mu \nu} G^{\mu \nu}= F_+ G_+ +F_- G_- \\
(F\,\widetilde{G})=&\frac{1}{2} \epsilon^{\mu \nu \rho \sigma}F_{\mu \nu}  G_{\rho \sigma}= -i(F_+ G_+ - F_- G_-)
\end{align}
Expanding the effective Lagrangian in (\ref{11}) we obtain
\begin{align}
\LL_\textup{eff}=&-
\frac{g_\textup{SM}^2 g_\textup{QFT}^2}{M^4}\Big[ (a_2-c_2+ b_3)(F_+ G_+)^2+(a_2-c_2-b_3)(F_- G_-)^2\nonumber\\
&+(a_1-c_1+ b_1+ b_2)(F_+ F_+)(G_+ G_+)+(a_1-c_1- b_1- b_2)(F_- F_-)(G_- G_-)\nn\\
&+(a_1-c_1+b_1 - b_2 )(F_+ F_+)(G_- G_-)+(a_1-c_1- b_1 + b_2)(F_- F_-)(G_+ G_+)\nonumber\\
&+ 2(a_2 + c_2)(F_+ G_+)(F_- G_-)\Big] \label{effectiveActionAFTER}
\end{align}
As already mentioned, in principle one has 16 different helicity configurations, however terms with an odd number of $\pm$ helicities identically vanish. We notice that in a non-chiral Lagrangian the coefficients $b_i$ must vanish due to parity symmetry.

Gauge invariance forbids terms with $M^0$ or $M^{-2}$ thus we expect that the contributions $A_{{{(0)}}}$ and $A_{{{(2)}}}$ must vanish in any helicity configuration.

\subsection{Comparison with the effective action}

In this section we compare the box amplitude in \ref{box_helicity} with the corresponding amplitude obtained from the effective action. We have noticed that the configurations $({+}{+}{+}{+})$ and $({-}{-}{-}{-})$ on the one end and $({-}{-}{+}{+})$ and $({+}{+}{-}{-})$ on the other end yield similar results with the same coefficients. This suggests that amplitudes are invariant under the exchange of positive and negative helicities, thus the effective action must be non-chiral. Therefore, all coefficients $b_i$ in \eqref{effectiveActionAFTER} vanish
\begin{equation}
b_1=0 \qquad b_2=0 \qquad b_3=0
\end{equation}
Using the same techniques as in \ref{box_helicity} we can compute the scattering amplitude from the effective action in the relevant helicity configurations obtaining
\begin{align}
A_\textup{  eff}^{++++}&={-} \frac{2 g_\textup{SM}^2 g_\textup{QFT}^2}{M^4} \Big[2(a_1{-}c_1)[p_1 p_2]^2 [p_3 p_4]^2{+} \\
& \hspace{3 cm} {+}(a_2{-}c_2)\big( [p_1 p_3]^2 [p_2 p_4]^2 {+}[p_1 p_4]^2 [p_2 p_3]^2\big )\Big] \nn \\
A_\textup{  eff}^{--++}&={-} 8 (2 a_1 + 4 c_1 + c_2)\frac{g_\textup{SM}^2 g_\textup{QFT}^2}{M^4} \frac{1}{4} \<p_1 p_2\>^2 [p_3 p_4]^2 \\
A_\textup{  eff}^{-+-+}&={-} 8 (a_2 + 2 c_1 + 2 c_2)\frac{g_\textup{SM}^2 g_\textup{QFT}^2}{M^4} \frac{1}{4} \<p_1 p_3\>^2 [p_2 p_4]^2 \\
A_\textup{  eff}^{-+++}&=0
\end{align}
Comparing these amplitudes with their counterparts in \ref{box_helicity}, we obtain the following coefficients for the effective action
\bea
a_1=-\frac{1}{90(4\pi)^2} = \frac{1}{2}a_2= \frac{4}{7}c_1= \frac{2}{7}c_2
\eea
which becomes
\begin{align}
\LL_\textup{eff}=
-\frac{g_\textup{SM}^2 g_\textup{QFT}^2}{90(4\pi)^2 M^4}\Big[&  (F \,F)(G\, G)+2 (F\, G)^2+ \frac{7}{4} (F\,\widetilde{F})(G\,\widetilde{G})+\frac{7}{2} (F\,\widetilde{G})^2\Big]
\end{align}
where appropriate traces are understood in the hidden and SM part. That agrees with similar results found by S. Marchesani under the supervision of Ya. Stanev \cite{Marchesani:2013}.

For comparison with the Euler-Heisenberg action, where only one gauge boson is present,
\begin{equation}
\LL_{E-H}=\LL_\textup{Kin}+\frac{e^4}{90(4\pi)^2 M^4} \left[(FF)^2 +\frac{7}{4}(F \widetilde{F})^2\right]
\end{equation}
we set $G=F$ and the new parameters are
\begin{equation}
a=\frac{a_1+a_2}{6}=-\frac{1}{90(4\pi)^2}\quad, \quad
c=\frac{c_1+c_2}{6}=-\frac{7}{360(4\pi)^2} \quad, \quad \frac{c}{a}=\frac{7}{4}
\end{equation}
which are in agreement with EH action.
The factor of 6 is due to the permutation symmetries of the interaction terms, in fact the 4-pt vertex $FFFF$ has an extra factor of 6=3!=4!/2!2! compared to $FFGG$.

\section{The multiscalar case}

In this appendix we generalize the analysis of section \ref{CoupledSectors} to the case of multiple scalar interactions linking the theories $T_1$  and $T_2$.

We start again from the two decoupled theories $T_1$ and $T_2$ and a set of scalar operators, $O_{i}(x)$, $i=1,\cdots, {M}$ belonging to $T_1$ and $\widetilde O_{\widetilde{i}}(x)$, ${a}=1,\cdots,\widetilde{M}$ belonging to $T_2$.

We define the partition function in the presence of sources
\be
Z(J_i(x))=\langle e^{i\int {d^4p\over (2\pi)^4}\sum_{i=1}^{M}J_i(p)O_i(-p)}\rangle
\label{a34}\ee
 and the Schwinger functional
 \be
 e^{iW(J_i)}={Z(J_i)\over Z(0)}
\label{a35} \ee
 which has as an expansion (in momentum space)
 \be
 W(J)={1\over 2} \int {d^4xd^4y}J_i(x)J_j(y)G_{ij}(x,y)+{1\over 3!}\int {d^4xd^4yd^4z}J_i(x)J_j(y)J_k(z)G_{ijk}(x,y,z)+\cdots
\label{a36}\ee

 Consider now the generating functional for the correlators of $O_i,\widetilde O_j$
 \be
Z(a_i,b_{\widetilde{i}})=\langle 0|e^{iS_{12}+i\int d^4x ~(\sum_{i=1}^M a_i(x) O_i(x)+\sum_{{\widetilde{i}}=1}^{\widetilde M}b_{\widetilde{i}}(x)\widetilde{O}_{\widetilde{i}}(x))}|0\rangle\sp  e^{iW(a_i,b_{\widetilde{i}})}={Z(a_i,b_{\widetilde{i}})\over Z(0,0)}
\label{a37} \ee
 with
 \be
 S_{12}=\int d^4x \sum_{i=1}^M\sum_{{\widetilde{i}}=1}^{\widetilde M} \lambda_{i{\widetilde{i}}}O_i(x)\widetilde O_{\widetilde{i}}(x)
\label{a38}\ee

If we ignore three- and higher-point  functions we have
\be
 e^{iW(a_i,b_j)}=N^{-1}\int \prod_{i=1}^M{\cal D} \phi_i\prod_{{\widetilde{i}}=1}^{\widetilde M}{\cal D} \widetilde \phi_{\widetilde{i}} ~e^{iS_{G}}
 \label{a39}\ee

\be
S_{G}=\int d^4x d^4y \left[\phi_i(x)G_{ij}^{-1}(x,y)\phi_j(y)+\widetilde\phi_i(x)\widetilde G_{{\widetilde{i}}{\widetilde{j}}}^{-1}(x,y)\widetilde \phi_{\widetilde{j}}(y)\right]+S_{12}+
\label{a40}\ee
$$
+\int d^4x ~(a_i(x) O_i(x)+b_{\widetilde{i}}(x) \widetilde O_{\widetilde{i}}(x))
$$
with
\be
\int d^4y~G_{ij}(x,y)G_{jk}^{-1}(y,z)=\delta_{ik}\delta^{(4)}(x-z)\sp G_{ij}(x,y)=\langle O_i(x)O_j(y)\rangle
\label{a41}\ee
and
\be
\int d^4y~\widetilde G_{{\widetilde{i}}{\widetilde{j}}}(x,y)\widetilde G_{{\widetilde{j}}{\widetilde{k}}}^{-1}(y,z)=\delta_{{\widetilde{i}}{\widetilde{k}}}\delta^{(4)}(x-z)\sp \widetilde G_{{\widetilde{i}}{\widetilde{j}}}(x,y)=\langle \widetilde O_{\widetilde{i}}(x)\widetilde O_{\widetilde{j}}(y)\rangle
\label{a41a}\ee
In momentum space and for translationally invariant theories
\be
G_{ij}(x,y)=\int {d^4p\over (2\pi)^4}e^{ip\cdot (x-y)}G_{ij}(p)\sp G_{ij}(p)G^{-1}_{jk}(p)=\delta_{ik}
\label{a42}\ee

In momentum space the action in (\ref{a37}) can be written as
\be
S={1\over 2}\int {d^4 p\over (2\pi)^4}\left[\phi_i(p)G^{-1}_{ij}(p)\phi_j(-p)+\widetilde\phi_{\widetilde{i}}(p)\widetilde G_{{\widetilde{i}}{\widetilde{j}}}^{-1}(p)\widetilde \phi_{\widetilde{j}}(-p)+2\l_{i{\widetilde{i}}}\phi_i(p)\widetilde \phi_{\widetilde{i}}(-p)+\right.
\label{a44}\ee
$$\left.
+2a_i(p)\phi_i(-p)+2b_{\widetilde{i}}(p)\widetilde \phi_{\widetilde{i}}(-p)\right]
$$
$$
={1\over 2}\int {d^4 p\over (2\pi)^4}\left[\left(\begin{matrix}\phi_i(p), &\phi_{\widetilde{i}}(p)\end{matrix}\right)\left(\begin{matrix}G_{ij}^{-1}(p)& -{\l_{i{\widetilde{j}}}}\\
-{\l_{{\widetilde{i}}j}}& \widetilde G_{{\widetilde{i}}{\widetilde{j}}}^{-1}(p)\end{matrix}\right)\left(\begin{matrix}\phi_j(-p)\\ \widetilde \phi_{\widetilde{j}}(-p)\end{matrix}\right)+2a_i(p)\phi_i(-p)+2b_{\widetilde{i}}(p)\widetilde \phi_{\widetilde{i}}(-p)\right]
$$
and the final Schwinger functional reads
\be
W(a_i,b_j)=
{1\over 2}\int {d^4 p\over (2\pi)^4}\left[\left(\begin{matrix}a_i(p), &b_{\widetilde{i}}(p)\end{matrix}\right)\left(\begin{matrix}G_{ij}^{-1}(p)& -{\l_{i{\widetilde{j}}}}\\
-{\l_{{\widetilde{i}}j}}& \widetilde G_{{\widetilde{i}}{\widetilde{j}}}^{-1}(p)\end{matrix}\right)^{-1}\left(\begin{matrix}a_j(-p)\\ b_{\widetilde{j}}(-p)\end{matrix}\right)\right]
\label{a45}\ee
with
\be
\left(\begin{matrix}G_{ij}^{-1}(p)& -{\l_{i{\widetilde{j}}}}\\
-{\l_{{\widetilde{i}}j}}& \widetilde G_{{\widetilde{i}}{\widetilde{j}}}^{-1}(p)\end{matrix}\right)^{-1}
=\left(\begin{matrix}[({1-\l \widetilde G\l G})^{-1}G]_{ij}& [({ 1-\l \widetilde G\l G})^{-1}G\l \widetilde G]_{i{\widetilde{j}}}\\
[({ 1-\widetilde G\l  G\l})^{-1}\widetilde G\l G]_{{\widetilde{i}}j}& [({ 1-\widetilde G\l  G\l})^{-1}\widetilde G]_{{\widetilde{i}}{\widetilde{j}}}\end{matrix}\right)
\label{a46}\ee

Therefore the new correlators for $\widetilde O_{\widetilde{i}}$ in momentum space are
\be
i\langle \widetilde O_{\widetilde{i}} \widetilde O_{\widetilde{j}}\rangle=(({ 1-\widetilde G\l  G\l})^{-1}\widetilde G)_{{\widetilde{i}}{\widetilde{j}}}
=\widetilde G_{{\widetilde{i}}{\widetilde{j}}}(p)+(({ 1-\widetilde G\l  G\l})^{-1}\widetilde G\l G\l)_{{\widetilde{i}}{\widetilde{j}}}
\label{a47}\ee
This formula generalizes the one in (\ref{a10}). We can now integrate in multiple scalars denoted by $\chi_{\widetilde{i}}$ and follow the procedure between eqn.~\ref{a16} and eqn.~\ref{a22} to obtain
\be
i \langle \widetilde O_{\widetilde{i}} \widetilde O_{\widetilde{j}}\rangle = (({ 1+ g \widetilde G   \widetilde K^{-1} g})^{-1}\widetilde G)_{{\widetilde{i}}{\widetilde{j}}}
\label{a48}\ee
This then leads to the identification for the emergent field matrix propagator
\be
\widetilde K_{\widetilde{i} \widetilde{j}}^{-1}(p) = - \frac{\left(\l G \l\right)_{\widetilde{i} \widetilde{j}}}{|g|^2} \, , \qquad \widetilde K_{\widetilde{i} \widetilde{j}}(p) = - |g|^2 \left(\l^{-1} G^{-1} \l^{-1}\right)_{\widetilde{i} \widetilde{j}}
\label{a49}\ee
as an immediate generalisation of~\ref{a22}. The couplings $g_{\widetilde{j}}$ do not affect much the previous discussion. The only way to obtain a hierarchy in the various elements of the matrix propagator is to consider a case where some of the $\lambda_{i \widetilde{j}}$ are hierarchically larger than the rest, since all the elements contain a trace over the hidden sector indices (and therefore all the hidden sector states contribute).

\section{Renormalization of axion coupling\label{apa}}

In this appendix we indicate how the quantum effects of SM (gauge) fields do not generate a mass for the axion in perturbation theory. We use EM as an example and do the calculation at one-loop.

Consider the theory defined by the (bare) action
\be
S_{PQ}=\int d^4x\left[-{1\over 4g^2}F_{\m\n}F^{\m\n}-{1\over 2}\partial_{\m}a\partial^{
\mu}a+{a\over M}\epsilon^{\m\n\rho\sigma}F_{\mu\nu}F_{\rho\sigma}\right]\sp F_{\m\n}\equiv \partial_{\m}A_{\n}-\partial_{\n}A_{\m}
\label{s}\ee
involving a scalar field $a$ and a U(1) gauge field $A_{\m}$. $\epsilon^{\m\n\rho\sigma}$ is the standard completely antisymmetric Levi-Civita tensor, with $\epsilon^{0123}=1$.

We vary the action
\be
\delta S=\int d^4x\left[-{1\over g^2}\partial_{\m}\delta A_{\nu}F^{\m\n}-\partial_{\m}\delta a \partial^{\mu}a+{\delta a\over M}
\epsilon^{\m\n\rho\sigma}F_{\mu\nu}F_{\rho\sigma}+{4a\over M}\epsilon^{\m\n\rho\sigma}\partial_{\mu}\delta A_{\nu}F_{\rho\sigma}\right]
\ee
and obtain the equations of motion by integrating by parts
\be
\partial_{\mu}F^{\m\n}+{4g^2\over M}\epsilon^{\m\n\rho\sigma}\partial_{\m}a F_{\rho\sigma}=0
\ee
\be
\partial_{\mu}\partial^{\mu}a+{1\over M}\epsilon^{\m\n\rho\sigma}F_{\mu\nu}F_{\rho\sigma}=0
\ee

We add the standard gauge fixing term ${1\over 2g^2\xi}(\partial_{\mu}A^{\mu})^2$ to the Lagrangian and derive the Feynman rules in momentum space.
Fourier transforming the action and scaling the gauge fields so that the kinetic term is simply normalized we obtain
\bea
S_{PQ}&=&\int {
d^4k\over (2\pi)^4}\left[-{1\over 2}A^{\m}(k)\left(k^2\eta_{\m\n}-k_{\m}k_{\nu}\right)A^{\nu}(-k)+{1\over 2\xi}A^{\m}(k)k_{\m}k_{\nu}A^{\nu}(-k)-{k^2\over 2}a(k)a(-k)\right]\nn\\
&&-{4g^2\over M}\int {d^4k_1d^4k_2\over (2\pi)^8}~\epsilon_{\m\n\rho\sigma}k_1^{\mu}k_2^{\rho}~a(-k_1-k_2)A^{\nu}(k_1)A^{\sigma}(k_2)
\eea
from which we read the photon propagator
\be
D_{\m\n}(k)={-i\over k^2+i\epsilon}\left[\eta_{\m\n}-(1-\xi){k_{\m}k_{\n}\over k^2}\right]
\ee
the scalar propagator
\be
S(k)={-i\over k^2+i\epsilon}
\ee
as well as the scalar-photon-photon vertex

\begin{equation}
\vcenter{\includegraphics[width=0.35\textwidth]{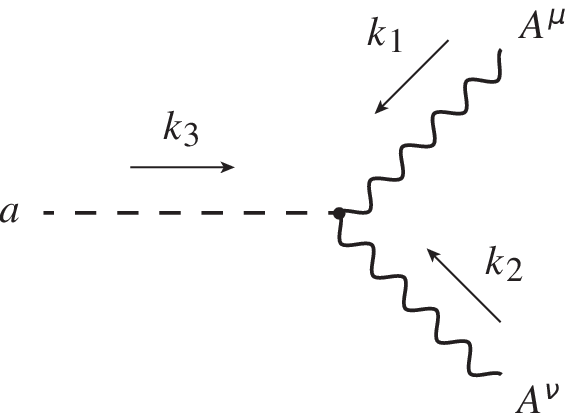}}
\hspace{-7cm} -i{4g^2\over M}\epsilon_{\mu\nu\rho\sigma}k_1^{\rho}k_2^{\sigma}
\end{equation}

\paragraph{1-loop renormalization of the scalar propagator.}

We now calculate the one-loop one-particle irreducible correction to the scalar propagator.
The relevant diagram is
\begin{equation}
~~~~~~~~~~~~~~~~~~~~~~~~I_{aa}~:~~~~~~~~~~~~~~~
\vcenter{\includegraphics[width=0.35\textwidth]{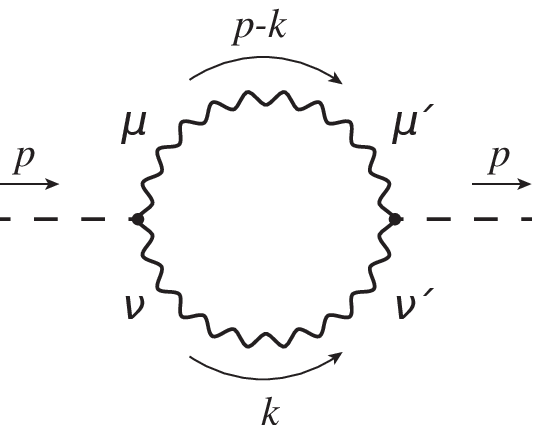}}
\end{equation}
and its amplitude reads
\bea
I_{aa}&=&\int{d^4k\over (2\pi)^4}\left(-i{4g^2\over M}\right)^2 \e_{\m\n\r\s}(p-k)^{\r}k^{\s}\e_{\m'\n'\r'\s'}(p-k)^{\r'}k^{\s'} \\
&&~~~~~~~~~~~\times
{(-i)\over k^2}\left[\eta_{\n\n'}-(1-\xi){k_{\n}k_{\n'}\over k^2}\right] {(-i)\over (p-k)^2} \left[\eta_{\m\m'}-(1-\xi){(p-k)_{\m}(p-k)_{\m'}\over (p-k)^2}\right]\nn\\
&=&2\left({4g^2\over M}\right)^2 \int{d^4k\over (2\pi)^4}{k^2(p-k)^2-(k\cdot (p-k))^2\over k^2(p-k)^2}
\eea
The $\xi$-gauge dependent part dropped out because of the antisymmetry of the $\e$ symbol.
Next we perform the following steps: (a) we introduce Feynman parameters and write $k^{-2}(k-p)^{-2} =\int^{1}_{0} dx (x(k-p)^2+(1-x)k^2)^{-2}$, (b) we shift $k\to k+xp$ in order to get the usual form of the denominator, (c) we rotate to Euclidean space by $k^0\to i k^4$ and we get
\bea
I_{aa}&=& -{3i\over 2}\left({4g^2\over M}\right)^2 p^2 \int_0^1dx\int{d^4k\over (2\pi)^4}{k^2\over  (k^2+(x-x^2)p^2)^2}
\eea
%
%
%
In order to proceed we introduce a cutoff $\Lambda$
\bea
I_{aa}&=&
-{3i\over 2}\left({4g^2\over M}\right)^2 p^2 \int_0^1dx\int{d\Omega_3 \over (2\pi)^4}\int_0^{\Lambda}k^3dk{k^2\over  (k^2+(x-x^2)p^2)^2}
\nn\\
&=&
-{3i\over 4}\left({4g^2\over M}\right)^2 p^2{\Omega_3 \over (2\pi)^4} \int_0^{\Lambda^2}{k^4 dk^2\over  (k^2+(x-x^2)p^2)^2}
\nn\\
&=&
-{3i\over 4}\left({4g^2\over M}\right)^2 p^2{\Omega_3 \over (2\pi)^4} \left[\Lambda^2+\int_0^1dx\left((x-x^2)p^2+2(x-x^2)p^2\log{(x-x^2)p^2\over \Lambda^2}+{\cal O}(\Lambda^{-2})\right)
\right]\nn\\
&=&-{3i\over 4}\left({4g^2\over M}\right)^2 p^2{\Omega_3 \over (2\pi)^4} \left[\Lambda^2-{7\over 18}p^2+{1\over 3}p^2\log{p^2\over \Lambda^2}+{\cal O}(\Lambda^{-2})\right]
\label{i3}\eea

The leading term is a quadratically divergent renormalization of the kinetic term of the $a$ field. In the messenger theory $\Lambda\sim M$.
The other terms generate higher derivative terms.
In particular note that there is no cutoff-dependent mass generated by the one-loop corrections.

The general expression for this renormalization to all loop orders involves
\be
\int {d^4 p\over (2\pi)^4} a(p)G(p) a(-p)
\ee
where $G(p)$ is the two-point function of the instanton density, in momentum space. It is well known, that in perturbation theory this correlation function
has no constant piece. For EM this is the whole story. However in a non-abelian theory strong YM dynamics generate a constant piece that is related to the topological susceptibility, \cite{Witten:1979vv, Veneziano:1979ec, DiVecchia:2017xpu,panago, DiVecchia:1981hh, DiVecchia:1981aev}. This will contribute a renormalization of the axion mass term as argued in the main text.

\section{One loop corrections through interactions}\label{Intloops}

\subsection{Path integrals with indices}\label{Indpath}

In order to set up the notations for the one-loop computation we will now consider path integrals where one has several fields combined into a vector. The latin indices $I,K = (1,2)$ take two values and denote the $\phi_1,  \phi_2$ fields. We work in Euclidean space, but is easy to extend these formulae into Lorentzian time by Wick rotating $\tau = - i t$.   Let us define the path integral
\be
Z[A, J]= \int \prod_{I=1,2} \mathcal{D} \phi_I e^{  - \half \int d^d x \phi_I A_{I K} \phi_K + \int d^d x J_I \phi_I} = \mathcal{C} \frac{e^{\int d^d x  \half J_I A^{-1}_{I K} J_K }}{\sqrt{\det A}}
\ee
where $A_{I K}$ is a matrix operator and thus the determinant is generically both a functional and a matrix one. 
To go to the non linear level one deforms this Gaussian theory with a generic potential $V(\phi_I)$
so that
\bea
Z &=& \mathcal{N} \int \prod_{I=1,2} \mathcal{D} \phi_I e^{  - \half \int d^d x \phi_I A_{I K} \phi_K + \int d^d x J_I \phi_I - V(\phi_I)}\, \nn \\
&=& \mathcal{N} e^{- \int d^d x V(\frac{\delta}{\delta J_I})} \int \prod_{I=1,2} \mathcal{D} \phi_I e^{ - \half  \int d^d x \phi_I A_{I K} \phi_K + \int d^d x J_I \phi_I } \nn \\
&=& \mathcal{N'} e^{- \int d^d x  V(\frac{\delta}{\delta J_I})} e^{  \half \int d^d x J_I A^{-1}_{I K} J_K } = \mathcal{N'} e^{ \half \int d^d x \frac{\delta}{\delta \phi_I} A_{I K}^{-1} \frac{\delta}{\delta \phi_K} } e^{ -  \int d^d x V(\phi_I) + J_I \phi_I} \vert_{\phi_I = 0}\, , \nn \\\label{EuclideanPathInt}
\eea
From this formal expression, upon taking derivatives one can construct all the Feynman diagrams of the theory and compute correlation functions once the field content of the theory is specified.

Note also that we do not write explicitly the overall normalization since there are functional determinants involved that need regularization and a more thorough study. For the correlation functions that we are interested in, such subtleties do not matter.

We will now study the one loop corrections to (\ref{d1}), (\ref{a9}), (\ref{a10})
 due to interaction terms (\ref{d2}). To read the one-loop corrected action (in a cutoff regulated fashion), we simply use the abstract expressions above to match any  Feynman diagram with the corresponding integral. The momenta running through the loops are defined up to a cutoff $\Lambda$ which in the main text is the messenger scale $M$. To fully renormalise the interacting theory, one can further follow a Wilsonian procedure using the path integral above. Here one makes a foliation in momentum space with a cutoff $\Lambda$ and integrates out the ``fast" modes fluctuating in a thin slice of momenta $(\Lambda- \delta \Lambda, \Lambda)$. These are the modes that run in the loop, but a general diagram has also dependence on the external momenta. This procedure for a general graph leads to the one-loop renormalization of the corresponding term in the effective action.

\subsection{Quartic vertex}\label{Quarticloop}

\begin{figure}[h]
\centering
\includegraphics[width=0.35\textwidth]{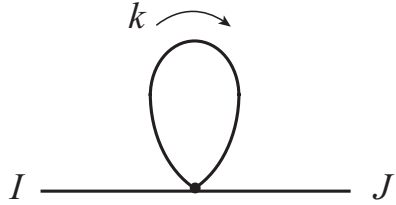}\\
\caption{The one-loop ``tadpole'' graph}
\label{axiondrop}
\end{figure}

The first example is a potential with a quartic vertex $V^{(4)}_{I J K L}(p)$. In this case one has the one loop diagram with two external lines (tadpole graph) given in fig.~\ref{axiondrop} that evaluates to
\be
 \Sigma_{I J}^{(\text{1-loop) (4)}}(p,-p) =   - \frac{1}{2} \int \frac{d^d k}{(2 \pi)^d} A^{-1}_{K L}(k) V^{(4)}_{I K L J}(k, p)
\label{1loop4v}
\ee

This expression is readily derived by expanding (\ref{EuclideanPathInt}) and keeping the quartic vertex term.
In our case the form of $A_{I K}^{-1}(p) = G_{I K}(p)$ is given in (\ref{a10}). We will now compute it for the simplest tensor structure for the quartic vertex, for which $V^{(4)}_{I K L J} = V_4 [\delta_{I K} \delta_{L J} +
\delta_{I L} \delta_{K J} ]$ (keep also in mind that our original 2-point function (\ref{a10}) is symmetric in the indices $I, J$). The relevant integrals can then be expanded in inverse powers of the cutoff using the formulae of Appendix~\ref{Cutoff}.

As an example if $G_{11}(p) \sim p^{2 \Delta_1 - 4}$ and $G_{22}(p) \sim p^{2 \Delta_2 - 4}$ we can express $\Sigma_{I J}^{(\text{1-loop) (4)}}$ in terms of the integral
$I(a,b,A,B)$ (\ref{mainint2}) as
\be
\mathcal{N} \begin{pmatrix}
I( 2 \Delta_2 - 1,  2(\Delta_1+\Delta_2) - 8 , - \lambda^2, 1) & \lambda I(2 (\Delta_1+\Delta_2) - 5,  2(\Delta_1+\Delta_2) - 8 , - \lambda^2, 1) \\
\lambda I( 2 (\Delta_1+\Delta_2) - 5, 2(\Delta_1+\Delta_2) - 8 , - \lambda^2, 1)&  I( 2 \Delta_1 - 1, 2\Delta_1+2 \Delta_2 - 8 , - \lambda^2, 1)
\end{pmatrix}
\label{B2}
\ee
with $\mathcal{N} = V_4 /16 \pi^2$. Expanding in powers of the cutoff $\Lambda$ using  (\ref{mainseries2}) one finds the matrix (\ref{CFT-CFT}) given in the main text. Other cases (such as a propagator with a mass gap) follow in a similar fashion.

\subsection{Cubic vertex}\label{Cubicloop}

\begin{figure}[h]
\centering
\includegraphics[width=0.4\textwidth]{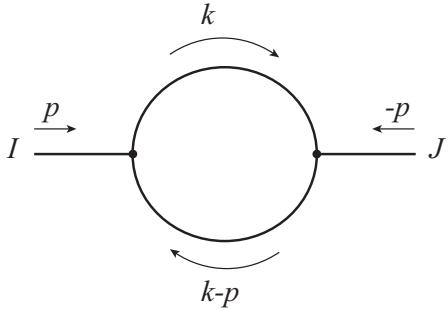}\\
\caption{The one-loop graph from cubic vertices.}
\label{axionloop}
\end{figure}

For the cubic interaction the one-loop correction is represented in fig.~\ref{axionloop} and evaluates to
\be
 \Sigma_{I J}^{(\text{1-loop) (3)}}(p,-p)  =   - \frac{1}{2} \int \frac{d^d k}{(2 \pi)^d} V^{(3)}_{I M L }(k, p) A^{-1}_{M M'}(k)  A^{-1}_{L L'}(k+p)  V^{(3)}_{M' L' J }(k, p)
\label{1loop3v}
\ee

We will be interested again in the case of a common vertex of strength $V_3$ so that one finds $3!=6$ possible terms that need to be summed over for each element. All the integrals can be re-expressed in terms of the following general form
\bea
I^{(\text{1-loop) (3)}} (a, b , c ;\, p, \Lambda)  & = &  V_3^2 \int^\Lambda d k k^{d-1} \frac{(k+p)^a k^b }{\left(1- \lambda^2 (k+p)^c \right) \left(1- \lambda^2 k^c \right)}\, , \nonumber \\
& = &  V_3^2 \int^\Lambda d k k^{d-1} \frac{(k+p)^a k^b }{\left(1- \lambda^2 (k+p)^c \right) \left(1- \lambda^2 k^c \right)}
\eea
in terms of a normalization prefactor and having indices $a,b,c$ that correspond to various momentum exponents.
Such integrals can be computed in a cutoff expansion as before using (\ref{mainint}), (\ref{mainint2}). To compute them, we use $(k+p)^2 =  k^2(1+ 2 p \cdot k / k^2 + p^2/k^2)$, expand for large $k$ and keep only the integrals with even powers of $k$, since integrals with odd powers of $k$ vanish (they are not rotationally invariant). If we just do power counting we immediately find the result in a similar fashion to the quartic case as an expansion in inverse powers of the cutoff. Due to consistency the first term will start again as in the case of the quartic vertex dictated by dimensional consistency (\ref{d3}) but now each power of the cutoff $\Lambda$ can be traded for a power of external $p$. There can be no odd powers of $p$ since these integrals are not rotationally invariant and vanish. This of course is a perturbative treatment for small $p/\Lambda$. When $p \sim \Lambda$ one needs to perform a more thorough study, since our low-energy approximation breaks down. To simplify the discussion in the main text we have picked a concrete interaction term of the form $\sim V_3  \phi_1^2 \phi_2 $. This is natural in case the ``hidden sector" is more strongly coupled than the visible SM and therefore graphs with more hidden sector fields appearing in internal propagators are contributing more.

\subsection{Cutoff regulated integrals}\label{Cutoff}

We will perform cutoff-regularization to the one loop integrals in order to keep power law divergences. The integrals we need are (for $d=4$, $a \geq 3$)
\be
I(a,A,B) = \int^\Lambda_0 d p \frac{p^a}{A p^2 + B}
\label{mainint}
\ee
The integral (\ref{mainint}) admits the following expansion for a large cut-off $\Lambda$
\be
I(a,A,B) \sim \frac{1}{2 A} \left[\frac{2 \Lambda^{a-1} }{(a-1) }+\frac{2 B \Lambda^{a-3} }{A(3 -a )
   }+O\left( \Lambda^{a-5}  \right)+\pi  \sec \left(\frac{\pi
   a}{2}\right) \left(\frac{A}{B}\right)^{\frac{1}{2}-\frac{a}{2}}\right] \label{mainseries}
\ee
One notices that when $a$ is an integer then the series breaks down and a corresponding term becomes logarithmically divergent. Some examples of this phenomenon are
\bea\label{quarticlogarithmic}
I(2,A,B) & = & \frac{\Lambda}{A}-\frac{\sqrt{B} \tan^{-1} \left(\frac{\sqrt{A} \Lambda}{\sqrt{B}}\right)}{A^{3/2}} \, , \nn \\
I(3,A,B) & = & \frac{\Lambda^2}{2 A}-\frac{B \log \left(A \Lambda^2+B\right)}{2 A^2} \, , \nn \\
I(4,A,B) & = & \frac{B^{3/2} \tan^{-1}\left(\frac{\sqrt{A} \Lambda}{\sqrt{B}}\right)}{A^{5/2}}+\frac{B
   \Lambda}{A^2}+\frac{\Lambda^3}{3 A} \, , \nn \\
I(5,A,B) & = & \frac{B^2 \log \left(A \Lambda^2+B\right)}{2 A^3}-\frac{B \Lambda^2}{2 A^2}+\frac{\Lambda^4}{4 A} \, , \nn \\
I(6,A,B) & = & -\frac{B^{5/2} \tan^{-1}\left(\frac{\sqrt{A} \Lambda}{\sqrt{B}}\right)}{A^{7/2}}+\frac{B^2
   \Lambda}{A^3}-\frac{B \Lambda^3}{3 A^2}+\frac{\Lambda^5}{5 A}
\eea
and so forth.
\\
\\
A more general integral will also be used
\be
I(a,b,A,B) = \int^\Lambda_0 d p \frac{p^a}{A p^b + B}
\label{mainint2}
\ee
with the large cut-off expansion
\be
I(a,b,A,B) \sim \frac{\Lambda^{a-3 b+1}}{A^3} \left(\frac{A^2 \Lambda^{2 b}}{a-b+1}-\frac{A B \Lambda^b}{a-2 b+1}+\frac{B^2}{a-3
   b+1}\right)+\frac{\pi   \csc \left(\frac{\pi  (a+1)}{b}\right)
   }{b B \left(\frac{A}{B}\right)^{\frac{a+1}{b}}} + ...
\label{mainseries2}
\ee
One again finds similarly logarithmically divergent terms in cases where $a + 1 - n b = m$ with $m,n$ integers.

\newpage


\end{document}